%% file: mainET_arXiv.tex
\documentclass[12pt]{article}
\usepackage[letterpaper, portrait, margin=1in]{geometry}
\usepackage{amsfonts,amsmath,amssymb,amsthm}
\usepackage[hidelinks]{hyperref}
\hypersetup{colorlinks,citecolor=blue,linkcolor=blue,urlcolor=blue}
\usepackage[hypcap]{caption}
\usepackage[utf8]{inputenc}
\usepackage[english]{babel}
\usepackage{graphicx}
\usepackage{natbib}
\usepackage{geometry}
\usepackage{booktabs}	
\usepackage{caption}
\usepackage{cases}
\usepackage{wrapfig}
\usepackage{float}
\usepackage{lscape}
\usepackage{rotating}
\usepackage{comment}
\usepackage{color}
\usepackage{multirow}
\usepackage{calc}
\usepackage[nameinlink]{cleveref}
\usepackage{setspace} 
\usepackage[shortlabels]{enumitem}
\usepackage{epstopdf}
\usepackage[final]{pdfpages}
\usepackage{csquotes}

\let\emptyset\varnothing

\usepackage[cal=boondox]{mathalfa}

\usepackage[bottom]{footmisc}
\def\signed #1{{\leavevmode\unskip\nobreak\hfil\penalty50\hskip2em
  \hbox{}\nobreak\hfil(#1)%
  \parfillskip=0pt \finalhyphendemerits=0 \endgraf}}

\newsavebox\mybox

\theoremstyle{plain}
\newtheorem{thm}{Theorem}
\newtheorem{lem}{Lemma}
\newtheorem{prop}{Proposition}

\newtheorem*{thm*}{Theorem}
\newtheorem{remark}{Remark}

\newtheorem*{fact*}{Fact}

\theoremstyle{definition}

\newtheorem{ex}{Example}

\theoremstyle{remark}

\newtheorem*{claim*}{Claim}

\title{Asylum Assignment and Burden-Sharing\thanks{This paper replaces and subsumes \citet{Capari2021}. We are grateful to Tayfun S\"{o}nmez, M. Utku \"{U}nver, M. Bumin Yenmez, David Delacretaz, Alex Teytelboym, and Kenzo Imamura  for their valuable feedback, as well as to participants at the Boston College dissertation workshop, the Nuffield postdoc seminar, Match-Up 2019, and the SAET Conference 2019. We are grateful to Yuan Gao for helpful research assistance. Gian Caspari acknowledges financial support from the Swiss National Science Foundation (SNSF) through a doctoral mobility stipend, facilitating his research visit to Oxford University during this project.}}

\author{Gian Caspari\thanks{Department of Market Design, ZEW — Leibniz Centre for European Economic Research, Mannheim 68161, Germany. Email: gian.caspari@zew.de.}
\and Manshu Khanna\thanks{Peking University HSBC Business School, Shenzhen 518055, China. Email: manshu@phbs.pku.edu.cn.}}

\begin{document}

\maketitle

\begin{abstract}	
We analyze the problem of matching asylum seekers to member states, incorporating wait times, preferences of asylum seekers, and the priorities, capacities, and burden-sharing commitments of member states. We identify a unique choice rule that addresses feasibility while balancing priorities and capacities. We examine the effects of both homogeneous and heterogeneous burden-sizes among asylum seekers on the matching process. Our main result shows that when all asylum seekers are treated as having identical burden-sizes, the asylum-seeker-proposing cumulative offer mechanism guarantees both stability and strategy-proofness. In contrast, when burden-sizes vary, there are scenarios where achieving stability or strategy-proofness is no longer possible.
\end{abstract}

\bigskip
\noindent\textbf{JEL Classification:} C62, C78, D47, J15

\bigskip
\noindent\textbf{Keywords:} Matching Theory, Market Design, Stability, Asylum Assignment

\newpage
\onehalfspacing

\section{Introduction}

When arriving in the European Union, an asylum seeker must submit an application for protection in a single member state. If successful, the person will be granted refugee status or subsidiary protection by the country that examined the asylum claim.
The responsible member state cannot be chosen freely. Under the Common European Asylum System (CEAS), the asylum seekers are required to lodge their application for protection in the country in which they initially arrive.\footnote{This rule is stated in Article 9 on page 62 of the \textit{Text of the agreement on Asylum and migration management regulation}. Document last accessed on 22 October 2024 at \url{https://www.europarl.europa.eu/meetdocs/2014_2019/plmrep/COMMITTEES/LIBE/DV/2024/02-14/06.RAMM_Asylumandmigrationmanagement_EN.pdf}.} This places a disproportionate burden on countries located at the border of the European Union. Its decentralized approach to asylum assignments leads to delays and disputes over responsibility, while its strict no-choice approach incentivizes asylum seekers to engage in illegal secondary movements to reach a more preferred member state \citep{EUREG2016}.

\subsection{An Alternative Asylum System }

A centralized asylum system provides an alternative to the current decentralized approach by considering both the preferences of asylum seekers and the priorities of member states. Under this system, member states retain control over eligibility determinations, while asylum seekers still apply for asylum at a single member state. Upon registration, asylum seekers would submit their preferences, and a centralized clearinghouse would then assign responsibility to a member state based on this information. Once a responsible member state is designated, asylum seekers must travel to and remain in that state while awaiting their asylum decision.

Unlike the current system, the centralized approach allows asylum seekers to express preferences for specific member states, giving them greater agency in the process. This ensures their preferences are factored into the assignment, making the system more responsive to their needs. Meanwhile, member states maintain control over eligibility determinations, allowing them to prioritize and regulate decisions according to their own standards. Importantly, the centralized system designates a responsible state before the asylum application is processed, avoiding the logistical challenge of relocating refugees after their status is determined.

The decentralized system often struggles to process claims in a timely manner, as required by the Charter of Fundamental Rights of the European Union \citep{beck2014application}.\footnote{The suggested timeline for a standard asylum claim is six months, and member states must inform applicants if the process takes longer. However, providing an estimated completion date does not imply a strict obligation to meet that timeframe \citep{beck2014application}.} The centralized system addresses this by allowing asylum seekers to express preferences for member states with shorter wait times, avoiding overburdened states and reducing delays. Recent evidence further underscores the importance of waiting times in shaping asylum seekers’ location decisions. \cite{farajzadeh2023optimizing} examine a sponsored parole program for Ukrainian refugees and find that waiting times significantly influence whether families accept an immediate offer of a sponsor in a less-preferred area, or instead opt to join a separate queue for a more desirable placement. Their results suggest that applicants will incur substantial additional waiting time to secure sponsorship in a preferred locality and that even small improvements in how waiting times are managed can have a major impact on overall matching efficiency and family well-being. These findings lend empirical support to the view that incorporating waiting-time considerations into asylum or parole matching systems is crucial to ensuring that location choices better reflect refugees’ actual preferences.

European regulators have also proposed burden-sharing quotas that set targets for the number of asylum applications each member state should process. As the \cite{EUProp2016} states, ``A corrective allocation mechanism should be established to ensure a fair sharing of responsibility between Member States [...] in situations when a Member State faces a disproportionate number of applications for international protection." We show that a centralized mechanism is particularly well-suited to facilitate the implementation of these burden-sharing quotas across the EU.

\subsection{Summary of Model and Analysis}

We formulate the asylum system as matching with contracts problem \citep{HatfieldMilgrom2005}, where a contract specifies an asylum seeker, a member state, and a wait time. Asylum seekers submit their preferences over combinations of wait times and member states. 
Following the CEAS, asylum seekers are restricted to a single application for protection, making it a many-to-one matching problem. 
An asylum seeker making an application can either represent an individual or a group of immediate family members applying for asylum at the same time. Therefore, each asylum seeker is given a burden-size.
Member states are required to make an asylum decision after the agreed-upon wait time ends. 
When scheduling asylum applications, member states are constrained by their bureaucratic capacities, representing the maximum number of asylum claims that can be processed in a given period. A member state's burden-sharing quota specifies the total amount of burden-size over asylum applications that have to be scheduled. Finally, to decide between different asylum seekers, each member state (strictly) ranks asylum seekers in terms of priority.\footnote{Member states' priority ranking criteria can be chosen freely as long as they don't interfere with article 14 of the European Convention on Human Rights \citep{jones2017international}, which prohibits discrimination on any grounds such as religion or race \citep{HumanR}. For example, in 2015, the former prime minister of Britain, David Cameron, announced the acceptance of up to 20,000 refugees from Syria, prioritizing  vulnerable children and orphans \citep{BBC2015}. Ties between asylum seekers with identical characteristics may be broken randomly whenever necessary.} 

A growing body of work employs mechanism design to account for states’ heterogeneous costs, refugees’ diverse preferences, and the political constraints over “who goes where.” In particular, \cite{hagen2022tradable, hagen2024optimal, hagen2024refugee} stress that effective relocation schemes must accommodate cross-country differences (e.g., population, GDP, attitudes toward immigration) and address informational hurdles (such as free-riding or misrepresentation). They also underscore that suitable matching or quota-trading frameworks hinge crucially on “distribution keys” specifying not only how many refugees each state should host but also why—these do not arise spontaneously, but rather from political bargaining, fairness doctrines, or prior agreements (e.g., the Dublin Regulation). In this paper, we treat burden-sharing quotas and bureaucratic capacities as exogenously fixed to focus on the matching design; however, \cite{moraga2021can} show that determining these parameters can itself be just as contentious as designing the matching mechanism.

We focus on finding a \textit{stable} and \textit{strategy-proof} mechanism for this problem. A stable mechanism avoids situations in which an asylum seeker prefers a different contract to the one she is assigned, and the member state specified in the more desirable contract is willing to handle her claim in the specified wait time. A strategy-proof mechanism (for asylum seekers) makes sure that misreporting preferences is not beneficial for asylum seekers, and thus it is their best interest to state their preferences truthfully.\footnote{We consider manipulations on the part of asylum seekers only. That is, we assume that member states cannot misreport priorities.}

We characterize a member state choice rule over contracts in \Cref{thm1}. This choice rule takes into account a member state's priority order, bureaucratic capacities, and burden-sharing quota, to make a selection from a given set of contracts. 
Given the member state choice rule, in \Cref{thm2} we show that: if asylum seekers are treated as having identical burden-sizes, the asylum-seeker-proposing cumulative offer mechanism \citep{HatfieldMilgrom2005} is stable and strategy-proof. However, when asylum seekers have different burden-sizes, Example 1 and Example 2 illustrate why both stability and strategy-proofness might not be achievable.

The proof of \Cref{thm2} builds upon the results developed in the literature on many-to-one matching markets with contracts. In particular, the member state choice rule (characterized in \Cref{thm1}) violates the substitutes condition and the law of aggregate demand \citep{kelso1982, HatfieldMilgrom2005, Hatfield2010}. However, for the case of homogenous burden-sizes, we show that a completion of the choice rule exists (\Cref{completion}(i)), as in  \citet{hatfield2016hidden}, that satisfies the substitutes condition if large burden-size asylum seekers are prioritized over smaller ones (\Cref{completion}(iii)). While the law of aggregate demand is satisfied if small burden-size asylum seekers are prioritized over larger ones (\Cref{completion}(iv)). The cumulative offer mechanism is stable and strategy-proof for identical burden-sizes because both requirements on the priority orderings are satisfied simultaneously.

We can obtain the same result by constructing an associated one-to-one market following \cite{kominers2016matching} to show that the cumulative offer mechanism is stable and strategy-proof for asylum seekers. We use \cite{hatfield2016hidden} because it helps us illustrate the lacking of cumulative offer mechanism in handling heterogeneous burden sizes. With heterogeneous burden sizes, standard guarantees such as stability, substitutability, and strategy-proofness can fail. When a member state’s priorities do not meet the large or small burden-size conditions, there may be no completion that satisfies substitutability or the law of aggregate demand (see \Cref{SubviolationCompletion} and \Cref{LADviolationCompletion}), and in general, no mechanism is both stable and strategy-proof (\Cref{thm3}). However, if priorities do satisfy large burden-size priority, the asylum-seeker-proposing cumulative-offer mechanism attains a stable outcome and is only non-obviously manipulable (\Cref{NOM for large burden size priority}), meaning that no profitable misreport strictly improves either an asylum seeker’s worst-case or best-case outcome, thereby ruling out obvious manipulations \citep[see][]{troyan2020obvious}.

\subsection{Related Literature}
The concept of managing refugee flows through organized systems has been discussed in the literature, starting with \citet{Schuck1997}, who proposed that each member state should bear a share of responsibility for temporary protection and permanent resettlement based on a quota system. \citet{moraga2014tradable} further developed this idea, proposing a system for trading quotas multilaterally without monetary exchange, aimed at resettling longstanding refugees and asylum seekers within the European Union. They were also the first to mention the possibility of combining a quota system with a matching mechanism, paving the way for subsequent models.

\citet{Jones2016,  jones2017international, Teytelboym} expanded on this by advocating for the use of matching mechanisms to enhance or replace existing resettlement practices. They differentiated between two levels of refugee matching: the global refugee match, operating at an international scale, and the local refugee match, which focuses on community-level integration within a country. In the international context, \citet{jones2017international, Teytelboym} suggested that the ``thickness" of the market allows refugee allocation to be effectively modeled as a standard school choice problem \citep{Balinski1999, Abdulkadiroglu2003}. Thus asylum seekers, whether individuals or families, can be treated equivalently in terms of burden-size, simplifying the allocation process. 

The local refugee match, however, presents more complex challenges \citep{jones2018local}. \citet{Delacretaz2016} introduced a framework that incorporates multidimensional knapsack constraints, accounting for the thinness of the market and the limitations of local resources. \citet{andersson2018dynamic} proposed a dynamic model that assigns refugees to localities based on types and locality-specific quotas. Additionally, \citet{andersson2020assigning} examined a market for allocating private housing to refugees, where landlords have preferences over the sizes of refugee families and the native languages they speak, highlighting the nuanced preferences that need to be considered in local resettlement. \citet{trapp2018placement} and \citet{dyntet} provided practical insights by assisting a U.S. resettlement agency in matching refugees to their initial placements, achieving improved employment outcomes.

Our work differs from previous studies by incorporating wait times into the global refugee match and addressing the differential burden-sizes across asylum seekers. Unlike past models, which assume the refugees’ status is determined prior to matching, our approach directly matches asylum seekers, allowing member states to retain control over the eligibility determination process. This conceptual shift necessitates the use of new framework based on matching with contracts model that must account for the strategic considerations over both member states and wait times.

Our work also relates to the literature on stable matchings with sizes, where complementarities arise from a combination of bureaucratic capacities and differential burden-sizes rather than hard capacity constraints \citep{dean2006unsplittable,mcdermid2010keeping, biro2014matching,yenmez2018college,nguyen2018near,delacretaz2019stability}. While the non-existence of stable allocations in matching with sizes is often due to rigid capacity limits, instability in our problem does not arise from the same, as burden-sharing quotas are not hard constraints. Nonetheless, \Cref{completion} reveals a tradeoff unique to matching problems involving agents of varying “sizes,” echoing well-known impossibility results: accommodating multi-unit agents (e.g., part-time versus full-time seats, day-care allocations, couples in the NRMP, or families in refugee resettlement) often forces a choice between competing notions of fairness or stability. In particular, as \citet{delacretaz2019stability} and \citet{Delacretaz2016} show, there is frequently no matching (or matching rule) that both avoids ‘waste’ and respects all capacity constraints when agents differ in their size of demand. Translating that tension to our setting, ensuring both substitutability and LAD with size-dependent priorities typically forces one to forgo at least one property in favor of the other.

Our work is also connected to the literature on matching with contracts, which has been explored in various contexts \citep{aygun2016dynamic,hassidim2017redesigning,yenmez2018college,Sonmez2013, Philipp2015}. Among these studies, our setup is most closely related to \citet{Philipp2015}. Of particular relevance is \citet{Philipp2015}, which examines the German entry-level labor market for lawyers, incorporating wait times as a contract term. While both models address wait times, our approach uniquely accounts for member state choice rules, the consideration of differential burden-sizes, and the inclusion of bureaucratic capacities, highlighting the unique complexities inherent in asylum assignment.

\section{Model and Definitions}
  
 An \textbf{asylum seeker matching problem} consists of

 \begin{enumerate}
 	\item a finite set of asylum seekers $A$,
 	
 	\item a finite set of member states $M$, 
 	
 	\item a finite set of wait times $W \subset \mathbb{R_+}$,

  \item a burden-size for each asylum seeker  $s:A\mapsto \mathbb{N}$,

    \item a list of burden-sharing quotas $q=(q_{m})_{m\in M}$ with $\sum_{m\in M} q_{m}\geq \sum_{a\in A} s(a)$.

 	\item a list of bureaucratic capacities $r=(r_{m}^{w})_{m\in  M, w\in W}$ with $\sum_{w\in W} r^w_m \geq q_m $ for all $m\in M$,
 	
 	\item a list of preference rankings $P=(\mathrel{P}_a)_{a\in A}$ over $M\times W$, and
 	
 	\item a list of priority orders $\pi=(\pi_m)_{m\in M}$ over $A$.
 \end{enumerate}

An asylum seeker's \textbf{burden-size} $s(a)$ specifies the amount of burden an asylum seeker's application consumes from a member state's burden-sharing quota. This allows an asylum maker applying on behalf of her immediate family differently from an asylum seeker applying as an individual. We say that an asylum seeker matching problem $\langle s,P\rangle$ specifies the \textbf{same burden-size} for every asylum seeker if $s(a)=s(a')$ for all $a, a' \in A$.

A member state's \textbf{burden-sharing quota} $q_m$ specifies the minimum number of applications a member state has to process. Since every asylum seeker has the right to submit an application, we require that the sum of burden-sharing quotas exceeds the sum of burden-sizes, that is, $\sum_{m\in M} q_{m}\geq \sum_{a\in A} s(a)$.

Each asylum seeker's application takes up one unit of \textbf{bureaucratic capacity}.  A member state $m$ can process at most $r_m^w\geq 0$ asylum seekers in time period $w$. Moreover, total bureaucratic capacities are sufficiently large to schedule any asylum seeker if a long enough time horizon $W$ is considered, that is, $\sum_{w\in W} r_{m}^{w}\geq |A|$ for all $m\in M$.

Each asylum seeker has a \textbf{preference ranking} $\mathrel{P}_a$ over wait-time-member-state combinations $M\times W$, with the corresponding weak preference written as $\mathrel{R}_a$.\footnote{$\mathrel{P}_a$ is a strict simple order, that is, a binary related that is \textit{transitive}, \textit{asymmetric}, and \textit{complete} (ranks everything except $(x,x)$ --- a contract with itself). The associated simple order $\mathrel{R}_a$ is \textit{transitive}, \textit{antisymmetric}, and \textit{strongly complete} (ranks everything).} Let $\mathcal{P}$ denote the set of all preference profiles.\footnote{One might assume that asylum seekers prefer lower wait times for the same member state. However, we allow for more general preferences to accommodate the concept of humanitarian visas. This means asylum seekers could opt for longer wait times outside the EU to allow enough time to apply for a humanitarian visa, which requires a designated member state at the time of application, and then travel there after the centralized match.}

Each member state has a \textbf{priority order} $\pi_m$ over the set of asylum seekers $A$,  indicating the order in which they would prefer to accept applicants.

We fix $A$, $M$, $W$, $q$, $r$, and $\pi$, denoting an asylum seeker matching problem by $\langle s,P\rangle$. 
To encompass wait times, we use a matching with contracts framework. 
A \textbf{contract} $x=(a,m,w)\in X=A\times M\times W$ specifies an asylum seeker $a\in A$, a member state $m\in M$, and a wait time $w\in W$. 
 Let $\mathcal{X}$ denote the set of all subsets of $X$.
Given a contract $x\in X$, let $a_x$ represent the asylum seeker, $m_x$ the member state, and $w_x$ the wait time specified in the contract. 
For some subset of contracts $X'\subseteq X$, let $X'_a=\{x\in X': a_x=a\}$ denote the set of contracts asylum seeker $a$ is part of, with equivalent notation for member states and wait times.
Let $A(X')=\{a\in A:a_x=a \text{ for some } x\in X'\}$ denote the set of asylum seekers specified in a subset of contracts $X'\subseteq X$, again equivalently defined for member states and wait times.
An \textbf{allocation} $Y\subseteq X$ is a set of contracts with $|Y_{a}| \leq 1$ for all $a \in A$,  and $|Y_{m}\cap Y_w|\leq r_{m}^w$ for all $m,w \in M\times W$.
Let $\mathcal{Y}$ denote the set of all allocations. Slightly abusing notation, we will use $\mathrel{P}_a$ for preferences over contracts and allocations.\footnote{For any $x,x'\in X$ we say $x\mathrel{P}_a x' \iff (m_{x},w_{x}) \mathrel{P}_a (m_{x'},w_{x'})$, and for any $Y,Y'\in \mathcal{Y}$ we say $Y\mathrel{P}_a Y' \iff Y_{a}\mathrel{P}_a Y'_{a}$.}

\section{Member State Choice Rule}
A choice rule for a member state makes a selection out of a set of contracts based on the member state's priority ordering, burden-sharing quota, and bureaucratic capacities. 
A \textbf{choice rule} $C_m:\mathcal{X} \mapsto \mathcal{X}$ associates with each subset of contracts $X'\in \mathcal{X}$ a subset of contracts $C_m(X')\in \mathcal{X}$ such that $C_m(X')\subseteq X'_m$.

We say that, an asylum seeker $a\in A(X')$ \textbf{qualifies for acceptance} under $X'\in \mathcal{X}$ if 
$$\sum_{\{a'\in A(C_m(X')): a'\pi_m a\}} s(a') < q_m.$$
For a given choice rule and a subset of contracts, consider the set of asylum seekers with at least one contract accepted. An asylum seeker qualifies for acceptance if the total burden-size of already accepted asylum seekers with a higher priority is strictly less than the member state's burden-sharing quota. Therefore, being “over quota” does not automatically disqualify an applicant. Instead, quotas serve as soft budget guidelines rather than absolute restrictions on assignment. 
 
We say that, an asylum seeker $a\in A(X')$ \textbf{qualifies for wait time} $w\in W$ under $X'\in \mathcal{X}$ if  $X'_a\cap X'_w\neq \emptyset$ and
$$|\{a' \in A(C_m(X')_{w}): a' \pi_m a\}|< r^{w}_m.$$

For a given choice rule, wait time, and a subset of contracts, 
consider the set of asylum seekers with a contract accepted for the relevant wait time. An asylum seeker qualifies for that wait time if the number of higher priority asylum seekers scheduled for that wait time is strictly less than the relevant bureaucratic capacity.

With this in mind, we propose a choice rule for member states that satisfies the following three properties:

\begin{itemize}

\item A choice rule $C_m$ satisfies \textbf{feasibility} if for all $X'\in \mathcal{X}$, for all $a\in A(X')$, and for all $w\in W$, we have $$|C_m(X')_a|\leq 1 \text{\;and\;} |C_m(X')_w|\leq r^w_m.$$

A choice rule is feasible if at most one contract is accepted per asylum seeker and the member state's bureaucratic capacities are not violated. This is unquestionably essential and is often assumed.

\item A choice rule $C_m$ satisfies \textbf{early filling} if for all $x\in C_m(X')$, and $x'\not\in C_m(X')$,
such that $a_{x}=a_{x'}=a$ and $w_{x'}<w_{x}$, the asylum seeker does not qualify for the lower wait time $w_{x'}$.

For a given subset of contracts, a choice rule satisfies early filling if for any accepted contract 
there does not exists a lower wait time contract for which there is still bureaucratic capacity left. 
This minimizes overall wait times at the member state. 

\item  A choice rule $C_m$ \textbf{respects member state priorities} if for all $X'\in \mathcal{X}$ an asylum seeker $a\in A(X')$ ends up with a contract if and only if she qualifies for acceptance and a wait time. 
 
 This property ensures that if an asylum envies another asylums seekers' contracts, then that asylum seeker must have a higher priority.

\end{itemize}

 Our main result in this section shows that there is a unique choice rule that satisfies feasibility, early filling, and respects member state priorities. The choice rule is straightforward and intuitive: asylum seekers are processed one by one in order of priority. Each is assigned the contract with the shortest waiting time that still has sufficient bureaucratic capacity. The process continues until either all asylum seekers have been matched or the burden-sharing quota is fully met, at which point no further contracts are accepted. 

\medskip
\noindent \textbf{Member state choice rule} $\hat{C}_m$ for a set of contract $X' \subseteq X$:

\noindent Step $k\geq 1$:
\begin{itemize}
	\item[1.] Let $X^{k-1}$ denote the set of accepted contracts with $X^0=\emptyset$, and let $Z^{k-1}=\{x\in X'_m:r^{w_x}_m>|X^{k-1}_{w_x}|\text{ and } a_x\not\in A(X^{k-1})\}$ denote the set of contracts specifying an asylum seeker not yet accepted with a wait time that has bureaucratic capacity left.\\
	 If either the burden-sharing quota is reached $\sum_{x\in X^{k-1}} s(a_x)\geq q_m$ or no acceptable contract is left $Z^{k-1}=\emptyset$ the algorithm ends and $\hat{C}_m(X')=X^{k-1}$. 
	
	\item[2.] Otherwise, determine the highest priority asylum seeker left $a^k$, defined as $a\in A(Z^{k-1})$ such that $a \mathrel{\pi} a'$ for all $a'\in  A(Z^{k-1})\setminus \{a\}$. \\
	Then, accept the lowest wait time contract  $x^k$ available for that asylum seeker, defined as $x\in Z_{a^{k}}^{k-1}$ such that $w_x<w_{x'}$ for all $x'\in Z^{k-1}_{a^{k}}\setminus \{x\}$.\\
	Adjust the set of accepted contracts $X^{k}=X^{k-1}\cup \{x^k\}$ and proceed to step $k+1$.
\end{itemize}

Hereafter, we will use $\hat{C}_m$ as the relevant choice rule, which is characterized in following result. Proof of \Cref{thm1} is relegated to \Cref{appendix}.

\begin{thm} 
\label{thm1}
A choice rule $C_m$ satisfies feasibility, early filling, and respects member state priorities if and only if it is $\hat{C}_m$.
\end{thm}

 \section{A Stable and Strategy-Proof Mechanism}
In this section, we introduce a mechanism with desirable properties designed to assign asylum seekers to member states and manage wait times. We will start with the  relevant definitions:

\begin{itemize}
    \item A \textbf{mechanism} is a function $\varphi$ that assigns every asylum seeker matching problem $\langle s,P \rangle$ an  allocation $\varphi(s,P) \in  \mathcal{Y}$. 

    \item A mechanism $\varphi$ is (pairwise) \textbf{stable} if for problem $\langle s,P \rangle$ if for every pair $a\in A$, $m\in M$ and contract $x\in X\setminus \varphi( s,P)$,

$$x \mathrel{P}_a \varphi( s,P) \implies x\not \in C_{m_x}(\varphi( s,P)\cup\{x\}). $$

A mechanism is considered (pairwise) stable if, whenever an asylum seeker prefers a contract over the one assigned to them by the mechanism, the member state involved would reject this alternative contract when evaluated alongside its existing set of assigned contracts. Thus, stability ensures fairness by preventing any party from benefiting at the expense of another.\footnote{We follow the matching with contracts tradition, defining stability via the absence of blocking pairs in agents’ choice functions (including priorities/constraints). While it is standard to require that no asylum-member state pair strictly prefers to deviate, this does not necessarily imply “no justified envy.” As \cite{romm2024stability} illustrate, stable allocations can still allow for higher-priority agents being left out while lower-priority agents are accepted. Nonetheless, focusing on pairwise blocking fully respects each side’s feasible choices by guaranteeing no local incentive to re-match and aligns with real-world decision rules.}

 \item A mechanism $\varphi$ is \textbf{strategy-proof} (for asylum seekers) for problem $\langle s,P \rangle$, if  for all $a\in A$, $ \hat{P}_a\in \mathcal{P}_a$, we have 
$$\varphi( s,P)\mathrel{R}_a \varphi( s,\hat{P}_a, {P}_{-a}).$$

A mechanism is strategy-proof if for every asylum seeker, revealing her true preferences is a weakly dominant strategy. 
\end{itemize}

\subsection{Homogenous Burden-size}

If asylum seekers are treated identically in terms of burden-sizes, then a stable and strategy-proof mechanism exists, namely, the asylum-seeker-proposing cumulative offer mechanism. We start by defining this mechanism:

\medskip
\noindent \textbf{Asylum seeker proposing cumulative offer mechanism} $\varphi^c$:

\textit{Step $k\geq 1$.}
\begin{center}
	\begin{itemize}
		\item[1.] Let $X^k$ denote the set of proposed contacts with $X^0=\emptyset$.\\
		A contract $x\in X^{k-1}$ is tentatively accepted if $x\in C_{m_{x}}(X^{k-1})$ and rejected otherwise.\\
		If there is any, let some asylum seeker $a$, with currently no contract tentatively accepted, propose her most preferred contract, following $P_a$,  among contracts that have not been proposed $x^k \in X\setminus X^k$. Then, set $X^{k}= X^{k-1}\cup \{x^k\}$. 
		\item[2.] Otherwise, the process terminates with $\varphi^c(s,P)=\bigcup_{m\in M}C_{m}(X^{k-1}_m)$.
	\end{itemize}
\end{center}

\begin{thm}
	\label{thm2}
	Suppose every member state is equipped with choice rule $\hat{C}_m$. Then the asylum-seeker-proposing cumulative offer mechanism is stable and strategy-proof for any problem $\langle s,P \rangle$ with identical burden-sizes across asylum seekers.
\end{thm}

A centralized system can assign asylum seekers to member states in a stable and (asylum-seeker) strategy-proof manner, provided families and individuals are treated equally in the burden-sharing quota during the assignment process. This requires identical treatment of applications from individuals and families. While the additional burden of supporting families cannot be factored in without compromising stability and strategy-proofness, imbalances can be addressed over time. Since asylum seeker matching is a recurring event, member states accepting more families could have their future burden-sharing commitments adjusted downward, balancing the overall burden.

Proving \Cref{thm2} requires conditions on the member state choice rule, which will together ensure that the above defined cumulative offer mechanism is stable and strategy-proof. 

\begin{itemize}
    \item A choice rule $C_{m}$ satisfies \textbf{substitutablility} if for all  $X'\subseteq X$, and $x,x'\in X\setminus X'$, we have
$$x\in C_{m}(X'\cup\{x,x'\}) \implies x\in C_{m}(X'\cup\{x\}).$$

A choice rule is substitutable if whenever a contract is chosen from a 
set of contracts, then the contract is also selected from any subset of contracts containing that contract. 

\item A choice rule $C_{m}$ satisfies \textbf{irrelevance of rejected contracts} if for all $X'\subseteq X$, and $x\in X\setminus X'$,
$$ x \not\in C_{m}(X' \cup \{x\}) \implies C_{m}(X' \cup \{x\})=C_{m}(X').$$

A choice rule satisfies irrelevance of rejected contracts if removing a contract that has not been chosen does not affect the set of chosen contracts.

\item A choice rule $C_{m}$ satisfies the \textbf{law of aggregate demand} if for all $X'\subseteq X$, and $x\in X\setminus X'$ we have
$$|C_{m}(X'\cup\{x\})|\geq |C_{m}(X')|.$$

A choice rule satisfies the law of aggregate demand if the set of chosen contracts weakly increases with the set of available contracts.

\end{itemize}

\cite{HatfieldMilgrom2005} and \cite{Aygun2013a} show that substitutability and irrelevance of rejected contracts are sufficient conditions on the choice rule for the cumulative offer mechanism to be stable. Additionally, combining these conditions with the law of aggregate demand ensures strategy-proofness of the cumulative offer mechanism.

\begin{prop}
	\label{IRC}
	The choice rule $\hat{C}_m$ satisfies the irrelevance of rejected contracts condition.
\end{prop}

However, wait times introduce complementarities regardless of the specified burden-sizes for asylum seekers.  Therefore, substitutability and law of aggregate demand are violated for the choice rule $\hat{C}_m$ even if  the asylum seeker matching problem $\langle s,P\rangle$ specifies the same burden-size for every asylum seeker. We present \Cref{choicefunctionviolatelad+s} to illustrate this point.

\begin{ex}[Choice rule violates substitutability and the law of aggregate demand]
	\label{choicefunctionviolatelad+s}
 Consider $A=\{a_1,a_2\}$, $M=\{m\}$, $W=\{w_l,w_h\}$. Let $s(a_1)=s(a_2)=1$,  $q_{m}=2$,  $r_{m}^{w_h}=r_{m}^{w_l}=1$, 
 	\begin{itemize}
 	\item[]$x_1=(a_1,m,w_l)$, $x_2=(a_1,m,w_h)$, 
 	\item[]$x_3=(a_2,m,w_l)$,  $x_4=(a_2,m,w_h)$, and
 	\item[]$\pi_{m}: a_1- a_2$.
 \end{itemize}
To see that $\hat{C}_m$ violates substitutability, consider $X'=\{x_2\}$. 
Noting that $x_4\in \hat{C}_m(X'\cup\{x_1,x_4\})=\{x_1,x_4\}$ while $x_4\not\in \hat{C}_m(X'\cup\{x_4\})=\{x_2\}$ reveals the failure of substitutability. Similarly, to see a violation of the law of aggregate demand, let $X'=\{x_2,x_3\}$. We then have $|\hat{C}_m(X')|=|\{x_2,x_3\}|>|\hat{C}_m(X'\cup \{x_1\})|=|\{x_1\}|$, confirming that $\hat{C}_m$ also violates the law of aggregate demand.
\end{ex}

\begin{remark}
    The choice rule in \Cref{choicefunctionviolatelad+s} also violates the weaker \textit{unilateral substitutes} condition \citep{Hatfield2010}, which together with the law of aggregate demand is sufficient for stability and strategy-proofness (for asylum seekers). A choice rule $C_{m}$ is unilateral substitutable if there does not exist $X'\subseteq X$ and $x,x'\in X\setminus X'$ such that $a_x\not\in A(X')$, and if $x \in C_{m}(X'\cup\{x,x'\})$ then $x\in C_{m}(X'\cup\{x\})$. We have a violation of unilateral substitutability as $a_{x_4}=a_2\not \in A(X')=\{a_1\}$.
\end{remark}

The literature has discussed several ways to relax the substitutability condition, as well as cases in which violations of substitutability and the law of aggregate demand are not harmful for stability and strategy-proofness \citep{Hatfield2010,kominers2016matching, hatfield2016hidden, hatfield2017stability}. In this paper, we rely on the result of \citet{hatfield2016hidden} which requires us to define a new property:

\begin{itemize}
    \item $C_{m}^{\prime}$ is a \textbf{completion} of a choice rule $C_{m}$, if for all $X'\subseteq X$, either $C_{m}^{\prime}(X')=C_{m}( X' )$, or there exist distinct contracts $x,x'\in C_{m}^{\prime}(X')$ that are associated with the same asylum seeker, that is, $a(x)=a(x')$.
    
    A choice rule completes another choice rule if they both choose the same set of contracts whenever the selected set of the completing choice rule is feasible.
\end{itemize}

\citet{hatfield2016hidden} show that if completion of a choice rule satisfies substitutability and the law of aggregate demand (together with the irrelevance of rejected contracts) then the asylum-seeker-proposing cumulative offer mechanism is stable and strategy-proof (for asylum seekers). The result holds even if the cumulative offer mechanism uses the original choice rule, which violates both properties.

 \begin{fact*}\citep{hatfield2016hidden}
If, for each $m\in M$, the choice function $C_m$ has a substitutable completion $C'_m$  that
satisfies the law of aggregate demand and the irrelevance of rejected contracts condition,
then the cumulative offer mechanism is stable and strategy-proof (for asylum seekers).
\end{fact*}

The following two restrictions on priority orderings will be useful before proceeding:

\begin{itemize}
    \item $\pi_m$ satisfies \textbf{small burden-size priority} if for all $a, a' \in A$, we have
$$a\pi_m a' \implies s(a) \leq s(a').$$

If an asylum seeker has higher priority than another asylum seeker, then she must have a weakly smaller burden-size.

\item $\pi_m$ satisfies \textbf{large burden-size priority} if for all $a, a' \in A$, we have
$$a\pi_m a' \implies s(a) \geq s(a').$$

If an asylum seeker has higher priority than another asylum seeker, then she must have a weakly larger burden-size.
\end{itemize}

We next describe a choice rule that is a completion of the choice rule characterized in \Cref{thm1}.

\medskip
\noindent \textbf{Completion of the member state choice rule} $\hat{C}'_m$:

\noindent Step $k\geq 1$:
\begin{itemize}
	\item[1.] Let $X^k$ denote the set of accepted contracts, with $X^0=\emptyset$.\\
	Similarly, let $Z^{k}=\{x\in X'_m:r^{w_x}_m>|X^{k}_{w_x}|\text{ and } x\not\in X^{k}\}$ denote the set of still acceptable contracts. \\
	If either the burden-sharing quota is reached $\sum_{x\in X^{k-1}} s(a_x)\geq q_m$, or no acceptable contract is left $Z^{k-1}=\emptyset$ the algorithm ends and $\hat{C}_m(X')=X^{k-1}$. 
	
	\item[2.] Otherwise, determine the highest priority asylum seeker left $a^k$, defined as $a\in A(Z^{k-1})$ such that $a \mathrel{\pi} a'$ for all $a'\in  A(Z^{k-1})\setminus \{a\}$. \\
	Then, accept the lowest wait time contract  $x^k$ available for that asylum seeker, defined as $x\in Z_{a^{k}}^{k-1}$, such that $w_x<w_{x'}$ for all $x'\in Z^{k}_{a^{k}}\setminus \{x\}$.\\
	Adjust the set of accepted contracts $X^{k}=X^{k-1}\cup \{x^k\}$ and proceed to step $k+1$.
\end{itemize}

The only difference of the completion $\hat{C}'_m$ relative to the original choice rule $\hat{C}_m$ is that an asylum seeker already holding a contract still remains in the race for more contracts. In particular, all contracts of a higher priority asylum seeker are accepted before all contracts with the identical wait time of a lower priority asylum seeker. Moreover, the completion $\hat{C}'_m$ satisfies the irrelevance of rejected contracts, is substitutable if the member states' priority order satisfies large burden-size priority, and satisfies the law of aggregate demand if small burden-size priority holds. 

\begin{prop}\label{completion}
The construction $\hat{C}'_m$ satisfies the following properties:
\begin{enumerate}[(i)]
   \item $\hat{C}'_m$ is a completion of $\hat{C}_m$. 
   \item $\hat{C}'_m$ satisfies irrelevance of rejected contracts.
   \item If $\pi_m$ satisfies large burden-size priority, then $\hat{C}'_m$ satisfies substitutability.
   \item If $\pi_m$ satisfies small burden-size priority, then $\hat{C}'_m$ satisfies the law of aggregate demand.
\end{enumerate}
\end{prop}

When the burden-size is identical across asylum seekers, the large and small burden-size conditions hold simultaneously. Therefore, these results collectively demonstrate that the cumulative offer mechanism is stable and strategy-proof (\Cref{thm2}). Proof of \Cref{completion} is relegated to \Cref{appendix}.

We revisit \Cref{choicefunctionviolatelad+s} to show that the completion addresses issues related to substitutability and the law of aggregate demand caused by bureaucratic capacities, provided the burden-size is homogeneous (\Cref{completionsublad}).

\begin{ex}[Completion \Cref{choicefunctionviolatelad+s} revisited]
	\label{completionsublad}
	
	 Recall the set-up from \Cref{choicefunctionviolatelad+s}:\\  $A=\{a_1,a_2\}$, $M=\{m\}$ $W=\{w_l,w_h\}$, $s(a_1)=s(a_2)=1$,  $q_{m}=2$,  $r_{m}^{w_h}=r_{m}^{w_l}=1$, and
	\begin{itemize}
		\item[]$x_1=(a_1,m_1,w_l)$, $x_2=(a_1,m_1,w_h)$, 
		\item[]$x_3=(a_2,m_1,w_l)$,  $x_4=(a_2,m_1,w_h)$,
		\item[]$\pi_{m}: a_1- a_2$.
	\end{itemize}	
	Revisiting the violation of substitutability for $X'=\{x_2\}$, $x_4$, and $x_1$. 
	We have that $\hat{C}'_{m}(X'\cup\{x_4,x_1\})=\{x_1,x_2\}$ as well as $\hat{C}'_{m}(X'\cup\{x_4\})=\{x_2\}$.
	Similarly, revisiting the law of aggregate demand violation for $X'=\{x_2,x_3\}$ and $x_1$, we have $|\hat{C}'_{m}(X')|=|\{x_2,x_3\}|=|\hat{C}'_{m}(X'\cup \{x_1\})|=|\{x_1,x_2\}|$.
\end{ex}

\subsection{Heterogenous Burden-size}

In the following two examples we illustrate the problem in using the cumulative offer mechanism when the burden-size is not identical across asylum seekers. If the member states' priority order does not satisfy large burden-size priority, \Cref{SubviolationCompletion} shows that there is no guarantee for the existence of a completion satisfying substitutability. If the member states' priority order does not satisfy small burden-size priority, \Cref{LADviolationCompletion} shows that there is no guarantee for the existence of a completion satisfying the law of aggregate demand.

\begin{ex}[No completion satisfying substitutability]
	\label{SubviolationCompletion}
	Consider $A=\{a_1,a_2,a_3\}$, $M=\{m\}$, $W=\{w_l,w_h\}$. Let $s(a_1)=1$, $s(a_2)=s(a_3)=2$, $q_{m}=2$,  $r_{m}^{w_l}=r_{m}^{w_h}=1$,
	\begin{itemize}
		\item[] $x_1=(a_1,m,w_l)$,  
		\item[] $x_2=(a_2,m,w_l)$,
		\item[] $x_3=(a_3,m,w_h)$, and
		\item[]$\pi_{m_1}: a_1 - a_2 - a_3$.
	\end{itemize}

	Note that priorities violate large burden-size priority.
	Moreover, for any $X'\subseteq X$, such that $|X'_a|\leq 1$ for all $a\in A$, any completion $\hat{C}'_m$ of $\hat{C}_m$ must choose the same set of contracts as $\hat{C}_m$, that is, $\hat{C}'_m(X')=\hat{C}_m(X')$.
	 We have that any completion must choose $\hat{C}'_m(\{x_2,x_3\})=\{x_2\}$ and $\hat{C}'_m(\{x_1,x_2,x_3\})=\{x_1,x_3\}$ which violates substitutability.

\end{ex}

\begin{remark}
    With \Cref{SubviolationCompletion}'s setup we can construct a counterexample for all sufficient conditions ensuring the existence of a stable mechanism. 
	We have that $\hat{C}_m(\{x_2,x_3\})=\{x_2\}$ and $\hat{C}_m(\{x_1,x_2,x_3\})=\{x_1,x_3\}$, a violation of \textit{bilateral substitutability} and therefore also of \textit{unilateral substitutability} as well as \textit{substitutability} \citep{HatfieldMilgrom2005,Hatfield2010}. 
	Note that $a_3$ gets a $w_h$-slot if $\{x_1,x_2,x_3\}$ are proposed but foregoes the slot if only  $\{x_2,x_3\}$ are proposed, so we cannot construct an \textit{associated one-to-one market} as in \cite{kominers2016matching}.  Finally, $x_2,x_3,x_1$ is an observable offer process, and thus we have a violation of \textit{observable substitutability} \citep{hatfield2017stability}, because $\hat{C}_m(\{x_2,x_3\})=\{x_2\}$ but $\hat{C}_m(\{x_1,x_2,x_3\})=\{x_1,x_3\}$.
\end{remark}

\begin{ex}[No completion satisfying the law of aggregate demand]
	\label{LADviolationCompletion}
	 Consider $A=\{a_1,a_2,a_3\}$, $M=\{m\}$, $W=\{w_l\}$. Let $s(a_1)=2$, $s(a_2)=s(a_3)=1$,  $q_{m}=2$,  $r_{m}^{w_l}=2$, and
	\begin{itemize}
		\item[] $x_1=(a_1,m,w_l)$,
		\item[]  $x_2=(a_2,m,w_l)$,
		\item[]  $x_3=(a_3,m,w_l)$,
		\item[] $\pi_{m}: a_1 - a_2 - a_3$.
	\end{itemize}
Note that priorities violate small burden-size priority.
Moreover, for any $X'\subseteq X$, such that $|X'_a|\leq 1$ for all $a\in A$, any completion $\hat{C}'_m$ of $\hat{C}_m$ must choose the same set of contracts as $\hat{C}_m$, that is, $\hat{C}'_m(X')=\hat{C}_m(X')$.
 It follows that, any completion must choose $\hat{C}'_m(\{x_2,x_3\})=\{x_2,x_3\}$ and $\hat{C}'_m(\{x_1,x_2,x_3\})=\{x_1\}$ which violates the law of aggregate demand.
\end{ex}

\begin{remark}
Note that $x_2,x_3,x_1$ is an observable offer process. Thus \Cref{LADviolationCompletion} also shows a violation of observable size monotonicity \citep{hatfield2017stability}. The same is true for \Cref{SubviolationCompletion}, which shows a violation of observable substitutability. Thus the weakest known conditions for existence of a stable and strategy-proof mechanism are violated.
\end{remark}

In general, when the burden-size varies across asylum seekers, there does not exist a stable mechanism (\Cref{notstableex}), nor can we ensure strategy-proofness to hold for a mechanism that selects a stable outcome, whenever one exists (\Cref{notstrategyproofex}).

\begin{thm}
	\label{thm3}
For a problem $\langle s,P \rangle$ with $|A| \geq 3$, $|M| \geq 4$ and any heterogenous burden-sizes, for some instance $\left[\left(\pi_m, \left(r_m^w\right)_{w \in W},q_m\right)_{m \in M}, W\right]$  of the asylum seeker matching problem,
\begin{enumerate}[(i)]
    \item a stable mechanism does not exist, and
    \item a strategy proof mechanism that selects a stable outcome (when possible) does not exist.
\end{enumerate}
\end{thm}

We present two examples to prove \Cref{thm3}.\footnote{Unlike previous results, for this result we need to construct burden-sharing quotas and bureaucratic capacities as well. Otherwise, sufficient capacities and quotas will allow for stable and strategy-proof mechanisms.} We will use $w_l<w_m<w_h$ to indicate low, medium and high wait time, respectively. 

\begin{ex}[No stable outcome]
	\label{notstableex}
	Let $W=\left\{w_l, w_m, w_h\right\}$. By heterogeneous burden-size, we can find three asylum seekers $a_1, a_2, a_3$ such that $s\left(a_1\right) \leq s\left(a_3\right)<s\left(a_2\right)$. Let $q_{m_1}=s\left(a_2\right), q_{m_2}=$ $q_{m_3}=1$. Let $q_{m_4}=\sum_{a \in A} s(a)$, $r_{m_4}^{w_l}=|A|$, and

 \begin{itemize}
\item[] $r_m^{w_n}=|A| \text { and } r_m^{w_h}=r_m^{w_l}=1 \text { for all } m \in\left\{m_1, m_2, m_3\right\},$
\item[] $P_{a_1}: x_1=\left(a_1, m_2, w_l\right)-x_2=\left(a_1, m_1, w_l\right)-\ldots, $
\item[] $P_{a_2}: x_3=\left(a_2, m_1, w_l\right)-x_4=\left(a_2, m_3, w_l\right)-\ldots, $
\item[] $P_{a_3}: x_5=\left(a_3, m_1, w_h\right)-x_6=\left(a_3, m_2, w_h\right)-\ldots, $
\item[] $P_{a_n}: x_{6+n}=\left(a_n, m_4, w_l\right)-\ldots \text { for all } n>3, $
\item[] $\pi_{m_1}: a_1-a_2-a_3, $
\item[] $\pi_{m_2}: a_3-a_2-a_1, $
\item[] $\pi_{m_3}: a_1-a_2-a_3, \text { and } $
\item[] $\pi_{m_4}: \ldots$
\end{itemize}

For any $a \notin\left\{a_1, a_2, a_3\right\}$, a stable outcome must match $a$ to $(a, m_4, w_l)$  as they like $m_4$ the most and $m_4$ has sufficient quota. For $a \in\left\{a_1, a_2, a_3\right\}$, we can verify that no stable outcome in this market: Since asylum seeker $a_1$ has the highest priority in member state $m_1=m_{x_2}$, in any stable allocation $a_1$ must end up with either $x_1$ or $x_2$. Suppose $a_1$ ends up with her top choice $x_1$ at member state $m_2=m_{x_1}$. In this case $a_2$ must get her top choice $x_3$, since she has the highest priority among the remaining asylum seekers at member state $m_1=m_{x_3}$, preventing $a_3$ from getting $x_5$ with member state $m_1=m_{x_5}$. In turn, $a_3$ forms a blocking pair with member state $m_2=m_{x_6}$ through contract $x_6$. Suppose $a_1$ gets $x_2$ instead. Then $a_2$ can no longer get her top choice $x_3$, due to insufficient bureaucratic capacity $r_m^{w_l}=1$ of member state $m_1=m_{x_3}$. On the other hand $a_3$ can get her first choice $x_5$ as there is sufficient bureaucratic capacity for the high wait time $r_m^{w_h}=1$ of member state $m_1=m_{x_5}$, while $a_2$ gets $x_4$ her second choice. But now $m_2$ has no asylum seeker assigned and $a_1$ forms a blocking pair with member state $m_2=m_{x_1}$ through contract $x_1$.  Thus, the non-existence of stable outcomes arises from a combination of burden-sizes and bureaucratic capacities.  
\end{ex}

\begin{ex}[No strategy-proof outcome]
	\label{notstrategyproofex}
	Let \(W=\{w_l,w_h\}\). Similarly by heterogeneous burden-size we can find \(a_1,a_2,a_3\) such that \(s(a_1)>s(a_2)\ge s(a_3)\). Let \( q_{m_1} = s(a_1) \), \( q_{m_2} = q_{m_3} = 1 \), \( r_m^{w_l} = 2 \) and \(r_{m}^{w_h}=\sum_{a\in A}s(a)\) for all \( m \in M \). Let \(q_{m_4}=\sum_{a\in A  }s(a)\), and 

\begin{itemize}
\item[] $P_{a_1} :  \ x_1 = (a_1, m_2, w_l) - x_2 = (a_1, m_1, w_l) - x_3 = (a_1, m_3, w_l) - \ldots, $
\item[] $P_{a_2} :  \ x_4 = (a_2, m_2, w_l) - x_5 = (a_2, m_1, w_l) - x_6 = (a_2, m_3, w_l), $
\item[] $P_{a_3} :  \ x_7 = (a_3, m_1, w_l) - x_8 = (a_3, m_2, w_l) - \ldots, $
\item[] $P_{a_n} :  \ x_{8+n}=(a_n,m_4,w_h)-\dots \text{ for all }n>3,$
\item[] $\pi_{m_1} :  \ a_1 - a_2 - a_3, $
\item[] $\pi_{m_2} :  \ a_3 - a_2 - a_1, $
\item[] $\pi_{m_3} :  \ a_1 - a_2 - a_3, \text { and } $
\item[] $\pi_{m_4}: \ldots$ 
\end{itemize}
For any stable mechanism \(\varphi\), we show it is not strategy-proof. \(\varphi(s,P)=Y_1 = \{x_2, x_6, x_8\} \cup \{x_{8+n} : n > 3, a_n \in A\}\) as \(Y_1\) is the unique stable allocation under \(P\). Now consider the misreport behavior of \(a_2\): If she reports \(\hat{P}_{a_2} : x_5 - x_6 - x_4\), under \(\left((P_a)_{a \ne a_2},\hat{P}_{a_2}\right)\) there are two stable allocations \(Y_1\) and \(Y_2 = \{x_1, x_5, x_7\} \cup \{x_{8+n} : n > 3, a_n \in A\}\). If \(\varphi(s,(P_a)_{a \ne a_2},\hat{P}_{a_2})=Y_2\), the manipulation is successful as \(\varphi(s,(P_a)_{a \ne a_2},\hat{P}_{a_2})=Y_2 \mathrel{P_{a_2}}\varphi(s,P)=Y_1\), so \(\varphi\) is not strategy-proof. If \(\varphi(s,(P_a)_{a \ne a_2},\hat{P}_{a_2})=Y_1\), consider the following misreport by \(a_1\): \(\hat{P}_{a_1}:x_1-x_3-x_2\). Under \(\left((P_a)_{a \notin \{a_1,a_2\}},\hat{P}_{a_2},\hat{P}_{a_1}\right)\) there is a unique stable allocation \(Y_2\), which implies \(\varphi(s,(P_a)_{a \notin \{a_1,a_2\}},\hat{P}_{a_2},\hat{P}_{a_1})=Y_2\). In this case, $a_1$ has a successful manipulation as \(\varphi(s,(P_a)_{a \notin \{a_1,a_2\}},\hat{P}_{a_2},\hat{P}_{a_1})=Y_2 \mathrel{P_{a_1}}Y_1=\varphi(s,(P_a)_{a \ne a_2},\hat{P}_{a_2})\). It follows that even if there exists a mechanism $\varphi$ that selects a stable outcome whenever one exists, strategy-proofness is violated for asylum seekers.	
\end{ex}

\subsubsection{Non-obvious manipulations}

With heterogeneous burden sizes, if priorities satisfy \emph{large burden-size priority}, the cumulative offer mechanism leads to a stable outcome.  
While the resulting mechanism is not strategy-proof, it satisfies the weaker criterion of \emph{non-obvious manipulability}.  
That is, even though there exist profitable deviations for some asylum seekers, these deviations do not strictly improve the ``worst-case'' or ``best-case'' outcome for the asylum seeker and thus fail to be an obvious manipulation \citep{troyan2020obvious}.

Recall that $P_a$ denotes
asylum seeker $a$'s \emph{true} ordinal preference ranking over wait-time-member-state combinations.  In line with \cite{troyan2020obvious}, we say a \emph{manipulation} $\hat{P}_a$ is profitable
if it yields a strictly better assignment for \emph{some} profile of preference reported by the other asylum seekers.  However, a mechanism $\varphi$ is \emph{not obviously manipulable (NOM)} if every profitable manipulation fails to strictly improve either the \emph{worst-case} \emph{or} the \emph{best-case} outcome for
the asylum seeker, when compared to truthful reporting.

Formally, for any given preference profile $P'_a \in \mathcal{P}_a$ let $ W_\varphi(a;P'_a) \;=\;\min\nolimits_{P_{-a}}\!\bigl(\varphi(s,P'_a,\,P_{-a})\bigr)$ denote the \emph{worst possible outcome} asylum seeker $a$ can ever obtain reporting $P_a$, and analogously let $B_\varphi(a;P'_a) \;=\;\max\nolimits_{P_{-a}}\!\bigl(\varphi(s,P'_a,\,P_{-a})\bigr)$ denote the \emph{best possible outcome} asylum seeker $a$ can ever obtain reporting $P'_a$ --- here $\min$/$\max$ is defined relative to asylum seeker $a$'s ordinal ranking $P_a$.

\begin{itemize}
\item A mechanism is
\textbf{not obviously manipulable (NOM)} if, whenever $a$ has a profitable lie $\hat{P}_a$, we do \emph{not} have
\[
  W_\varphi(a;\hat{P}_a) \;P_a\; W_\varphi(a;P_a)
  \quad\text{nor}\quad
  B_\varphi(a;\hat{P}_a) \;P_a\; B_\varphi(a;P_a).
\]
Otherwise, if some profitable deviation yields a strictly better worst-case
or strictly better best-case assignment, we say the mechanism is
\textbf{obviously manipulable}.
\end{itemize}

\begin{prop}
    \label{NOM for large burden size priority}
   Suppose every member state is equipped with choice rule $\hat{C}_m$ and its priority order $\pi_m$ satisfies large burden-size priority. Then the asylum-seeker-proposing cumulative offer mechanism is stable and non-obviously manipulable for any problem $\langle s,P \rangle $.
\end{prop}

\section{Discussion}

\paragraph{Large Quotas and Capacities.}
The EU’s legal architecture already provides a baseline for our assumption that every asylum seeker will ultimately be processed: under the Dublin III Regulation, a single Member State must examine each claim, and the EU Pact on Migration and Asylum aims to introduce a compulsory solidarity mechanism that shares the burden among states more equally. Hence, every asylum seeker must, at least in principle, be processed somewhere. In our model, we presume the existence of a solidarity mechanism, and this de facto obligation is captured by positing quotas whose sum surpasses the total burden.
Furthermore, we allow cumulative bureaucratic capacity—interpretable as the number of available case officers—to grow with the time horizon, mirroring how administrations hire staff or redeploy resources when inflows rise. Should a particular state be temporarily unable to process an applicant, the situation can be represented as assignment to a “waiting-room” state (a fictitious member with a very large quota), so the overall accounting identity is preserved even if the real burden-sharing formula remains politically contested. This theoretical device is not merely a mathematical convenience: in practice, cases are often rolled over to subsequent months or years until new capacity comes online, and German administrative practice illustrates that blanket refusals for capacity reasons would face significant legal hurdles, reinforcing our no-rejection premise.\footnote{See, for example, the European Council on Refugees and Exiles (ECRE) report on Germany’s regular asylum procedure:
  \url{https://asylumineurope.org/reports/country/germany/asylum-procedure/procedures/regular-procedure}.}

\paragraph{Burden-Size and Bureaucratic Capacity.}
We treat burden-size as an individual characteristic of an asylum seeker (e.g.\ single applicant vs.\ family). One could argue that the effective burden also depends on the receiving member state's composition of staff and its attitudes: a state might, for instance, consider certain subgroups less burdensome than others. While one can easily extend our model to be state-specific in the function $s_m(a)$, we abstract away from such heterogeneity for simplicity. As for the concern that families might require a larger share of the \emph{bureaucratic capacity} than single applicants, a more fine-grained model could differentiate additional capacity costs based on family size or other special needs. Here, we assume each application is processed by exactly one ``slot,'' mirroring real-life administrative practice where each \emph{case} is assigned to a single case officer (irrespective of the family size).\footnote{For details on maintaining family unity beyond the standard family size, see Article 23 (“Maintaining family unity”) of Regulation (EU) 2024/1347 of the European Parliament and of the Council, available at: \url{https://eur-lex.europa.eu/legal-content/EN/TXT/?uri=OJ:L_202401347}.} These representations of quota and capacity are consistent with the approach taken in other matching-with-contracts models, yet we acknowledge potential variations in how bureaucratic workloads could be subdivided.

\paragraph{Homogenous vs.\ Heterogenous Burden-Size.}
When we speak of burden we mean the share of work that a member state is expected to shoulder under whatever burden-sharing arrangement is assumed to be in force, not the full, real-world load that may ultimately fall on that state. The simplest yardstick—used in our model—is a head-count of the asylum applications the administration must process, so a family that files a single joint application (as in Germany) is counted once and imposes the same formal burden as a single adult. Treating every case as identical obviously glosses over the rich variety of costs that follow from differences in family size, health status, housing needs, or labour-market prospects; countries even diverge on whether they tally each family member separately (“family of four = four units”) or regard the file as one case. The situation is further complicated by the fact that not all asylum seekers are ultimately granted refugee status, leading to variation in the duration of required support. We therefore frame the equal-burden assumption as a clean baseline that lets us highlight how a stable, strategy-proof matching mechanism operates.

Future extensions could move beyond the equal-burden baseline by weighting cases according to their expected resource demands—larger families, applicants with significant healthcare needs, or those requiring specialised housing would count for more capacity than a single adult—thus preserving the core matching logic while better reflecting real-world heterogeneity. A complementary line of inquiry is to relax strict stability or capacity constraints, following evidence from \citet{nguyen2021stability} that modest violations of these conditions can deliver sizable efficiency gains. Allowing temporary capacity overflows or giving states flexible quotas might therefore enable assignment mechanisms to accommodate diverse case profiles without abandoning their strategy-proof foundations, offering policymakers an additional lever for distributing the long-term responsibilities of receiving differently sized asylum-seeker cohorts.

\section{Conclusion}

We presented an alternative to the current decentralized asylum assignment to effectively match asylum seekers to member states, taking into account the complex interplay of preferences, priorities, capacities, and burden-sharing commitments.  We demonstrate that a centralized mechanism can ensure stability and strategy-proofness under conditions of equal burden-sizes. However, when the burden-sizes differ, the feasibility of achieving these desirable properties is challenged.

While the homogeneous burden-size assumption simplifies the problem and leads to a desirable solution, it might not reflect the real-world complexities of asylum assignment, where some member states may face disproportionately higher numbers of asylum seekers with families. To address the challenges posed by heterogeneous burden-sizes, one may consider adjusting burden-sharing quotas over time by rewarding member states that accept a disproportionate number of families by lowering their quotas in subsequent assignment rounds. Such an approach could help mitigate the long-term effects of burden imbalances. 

In terms of practicality, our proposed mechanism can be implemented at regular intervals as a recurring process, allowing the batch of newly arrived asylum seekers to incorporate their preferences into the asylum assignment process. This approach is similar to Japan's centralized daycare center allocation, which is not based on a first-come, first-served basis. Instead, slots are assigned annually in April via a centralized algorithm that considers family preferences and prioritizes children from low-income households, single-parent families, or those with guardians facing health challenges \citep{sun2023daycare}. German municipalities also allocate daycare places at regular intervals.\footnote{See \url{https://kitamatch.com/}.} Another example of this type of matching mechanism is Singapore's Build-To-Order (BTO) system for public housing, where flats are allocated monthly based on buyers' preferences and a priority system. While asylum seekers arrive continuously, processing applications in batches at regular intervals could significantly improve the current decentralized system.

\pagebreak

\include{library2.bib}
\bibliographystyle{aer}
\bibliography{library2}

\pagebreak
\appendix

\section{Mathematical Appendix and Proofs}\label{appendix}

\subsection*{\Cref{thm1}}		
\begin{proof}

	We start with the \textbf{if} direction.\\
	\textit{Feasibility}: Suppose feasibility is violated and consider the first step $k$ in which some asylum seeker $a^k$ gets her second contract. That is, $a^k\in A(Z^{k-1})$ such that $a^k \pi a'$ for all $a'\in  A(Z^{k-1})\setminus \{a^k\}$. Note that $Z^{k-1}=\{x\in X'_m:r^{w_x}_m>|X^{k-1}_{w_x}|\text{ and } a_x\not\in A(X^{k-1})\}$ but by assumption we have $a^k\in  A(X^{k-1})$ as one contract has already been accepted: leading to a contradiction. 
	Similarly, consider the first step $k$, in which $|X^{k-1}_w|= r^w_m$ and $w_{x^k}=w$, again we get a contradiction with $a^k\in A(Z^{k-1})$ as $r^{w_x}_m=|X^{k-1}_{w_x}|$.

	\noindent \textit{Early filling}: Consider some asylum seeker $a^k$ who got assigned contract $x^k$ at step $k$.
	Note that the algorithm chooses the lowest contract in $Z^{k-1}_{a^k}$, hence for $x'\in X_{a^k}$ with $w_{x'}<w_{x^k}$ we have that $r^{w_{x'}}_m =|X^{k-1}_{w_x'}|$. Since all asylum seekers in $X^{k-1}$ have higher priority than $x^k$, it follows directly that $|\{a' \in A(X^{k-1}_{w_x'}): a' \pi a\}|=|\{a' \in A(C_m(X')_{w_{x'}}): a' \pi a\}|= r^{w_x}_m$, and thus early filling is satisfied. 
	
	\noindent \textit{Respecting member state priorities}:
	 Consider $a^k$ accepted at step $k$. We have that $\sum_{x\in X^{k-1}} s(a_x)< q_m$ and since $a^k$ has higher priority than all remaining asylum seekers, we have 
	$\sum_{x\in X^{k-1}} s(a_x)=\sum_{\{a'\in A(C_m(X')): a'\pi_m a\}} s(a') < q_m$.
	
	\noindent Similarly, if the algorithm ends at step $k$, we have that $\sum_{x\in X^{k-1}} s(a_x)\geq q_m$ and as $a^k$ has lower priority than all previously chosen asylum seekers, we have 
	$\sum_{x\in X^{k-1}} s(a_x)=\sum_{\{a'\in A(C_m(X')): a'\pi_m a\}} s(a') \geq q_m$.
	
	\noindent In step $k$, only contracts $Z^{k-1}=\{x\in X'_m:r^{w_x}_m>|X^{k}_{w_x}|\text{ and } a_x\not\in A(X^{k})\}$ are considered.
	Take any unassigned asylum seekers $a\in A(X')\setminus \{a^1,\dotso, a^{k-1}\}$: clearly any contract $x\in X_a'\setminus Z^{k-1}$ does not qualify for a wait time as the bureaucratic capacity is already occupied with higher priority asylum seekers,  $|X^{k-1}_{w_x}|=|\{a' \in A(X^{k-1}_{w_x}): a' \pi a\}|= |\{a' \in A(C_m(X')_{w_x}): a' \pi a\}|=r^{w_x}_m$.
	Similarly, if $a^k$ is accepted in step $k$ then $x^k\in Z^{k-1}$ and hence $|X^{k-1}_{w_x}|=|\{a' \in A(X^{k-1}_{w_x}): a' \pi a^k\}|= |\{a' \in A(C_m(X')_{w_x}): a' \pi a^k\}|\leq r^{w_x}_m$.
	
	\noindent Hence, almost by construction, an asylum seeker is accepted if and only if she qualifies for acceptance and a wait time.

	\noindent We show the \textbf{only if} direction, proceeding by induction.
	Suppose the described algorithm for determining $\hat{C}_m(X')$ stops after $k$ steps, and therefore $\hat{C}_m(X')=X^{k-1}$. 
	
	\noindent\textbf{Base step}. If the algorithm stops at step $1$ then $\hat{C}_m(X')=C_m(X')=X^0$; otherwise, $X^1\subseteq C_m(X')$.

	\noindent Suppose the algorithm stops at step $1$.\\
	We have $\sum_{x\in X^{0}} s(a_x)=\sum_{x\in \emptyset} s(a_x)=0\geq q_m$. It follows that no asylum seeker \textit{qualifies for acceptance}, since for all $a\in A(X')$ we have $\sum_{\{a'\in A(C_m(X')): a'\pi_m a\}} s(a') \geq q_m$. 
	It follows that $\hat{C}_m(X')=C_m(X')=X^0=\emptyset$ since $C_m(X')$ satisfies \textit{respecting member state priorities}.
	
	\noindent Suppose the algorithm does not stop at step $1$.\\
	No contract in $X'\setminus Z^0$ can ever be accepted without violating \textit{feasibility}, hence the only relevant asylum seekers are in the set $A(Z^0)$.
	Moreover, as the algorithm did not stop, we have $\sum_{x\in X^{0}} s(a_x)=\sum_{x\in \emptyset} s(a_x)=0< q_m$ and $Z^{0}\neq \emptyset$.
	As $a^1$ is the highest priority asylum seeker, we have that $\sum_{\{a'\in A(C_m(X')): a'\pi_m a\}} s(a')< q_m$ and hence $a^1$ \textit{qualifies for acceptance}. 
	Similarly, $|\{a' \in A(C_m(X')_{w_x}): a' \pi a^1\}|\leq r^{w_x}_m$ holds for all $x\in Z^0_{a^1}$ and therefore at least one contract in $Z^0_{a^1}$ must be accepted since $a^1$ also \textit{qualifies for a wait time}.
	Due to \textit{early filling}, the lowest available wait time contract must be accepted, which in this case is $x^1$, that is $x^1\in C_m(X')$ and therefore $X^1=\{x^1\}\subseteq C_m(X')$. In other words, the contract accepted under the described choice rule --- during step $1$ of the described algorithm, that is, $X^1\subseteq \hat{C}_m(X')$ --- must also be accepted under any other choice rule satisfying the described axioms $X^1\subseteq C_m(X')$.

	\noindent\textbf{Induction step}. We assume that if the algorithm stops at step $k-1$ then $\hat{C}_m(X')=C_m(X')=X^{k-2}$, otherwise $X^{k-1}\subseteq  C_m(X')$.
	Given that, if the algorithm stopped at step $k$ then $\hat{C}_m(X')=C_m(X')=X^{k-1}$, otherwise $X^{k}\subseteq  C_m(X')$.
	
	
	\noindent Note that if the algorithm stops at step $k-1$ then $\hat{C}_m(X')=C_m(X')=X^{k-2}$ by the induction assumption. 
	
	\noindent Suppose the algorithm stops at step $k$.\\
	We have $\sum_{x\in X^{k-1}} s(a_x)\geq q_m$. It follows that no asylum seeker \textit{qualifies for acceptance}, because due to the induction assumption for all remaining asylum seekers $a\in A(X^{k-1})$ we have 
	$\sum_{\{a'\in A(C_m(X')): a'\pi_m a\}} s(a') \geq q_m$ since $X^{k-1}\subseteq C_m(X')$. 
	It follows that $\hat{C}_m(X')=C_m(X')=X^{k-1}$ as $C_m(X')$ satisfies \textit{respecting member state priorities}.
	
	\noindent Suppose the algorithm does not stop at step $k$.\\
	No contract in $X'\setminus Z^{k-1}$ can ever be accepted without violating \textit{feasibility}. Hence, the only relevant asylum seekers are $A(Z^{k-1})$.
	As the algorithm did not stop, we have $\sum_{x\in X^{k-1}} s(a_x)< q_m$ and $Z^{k-1}\neq \emptyset$.
	As $a^k$ is the highest priority asylum seeker, we have that $\sum_{\{a'\in A(C_m(X')): a'\pi_m a\}} s(a')< q_m$ and thus $a^k$ \textit{qualifies for acceptance}. 
	Similarly,  $|\{a' \in A(C_m(X')_{w_x}): a' \pi a^k\}|\leq r^{w_x}_m$ holds for all $x\in Z^{k-1}_{a^k}$ and hence at least one contract in $Z^{k-1}_{a^k}$ must be accepted as $a^k$ also \textit{qualifies for a wait time}.
	Due to \textit{early filling} the lowest available wait time contract must be accepted, which in this case is $x^k$, that is $x^k\in C_m(X')$ and therefore together with the induction assumption $X^k=X^{k-1}\cup \{x^k\}\subseteq C_m(X')$.
	
	\noindent As the algorithm describing $\hat{C}_m(X')$ ends after a finite number of steps, we have $\hat{C}_m(X')=C_m(X')$.
	
\end{proof}

\subsection*{\Cref{IRC}}		

\begin{proof}
	\noindent Consider $X'$ and $X'\cup \{x^*\}$ for some $x^*\in X\setminus X'$. 
	We refer to the relevant sets during each step of the algorithm for $C_m$ as $X^{k\prime\prime}$, $Z^{k\prime\prime}$, and so on under the former ($C_m(X')$) and $X^{k\prime}$, $Z^{k\prime}$, and so on under the latter ($C_m(X'\cup \{x^*\})$). 
	
	\noindent Given that $x^*\not \in C_m(X'\cup \{x^*\})$, we want to show that $C_m^{\prime}(X'\cup \{x^*\})=C_m(X')$. We proceed by induction.
	
	\noindent \textbf{Base step}: We assume that $x^*\not \in C_m(X'\cup \{x^*\})$. If the former algorithm ($C_m(X')$) stops at step $1$ we have $C_m(X'\cup \{x^*\})=C_m(X')$ and $X^{1\prime}=X^{1 \prime \prime}$ otherwise.
	
	
	\noindent Suppose the former algorithm stops at step $1$.\\
	Case 1: If $\sum_{x\in X^{0\prime \prime}} s(a_x)\geq q_m$ then $\sum_{x\in X^{0 \prime}} s(a_x)\geq q_m$ as $X^{0\prime \prime}=X^{0\prime}=\emptyset$ and therefore $C_m(X'\cup \{x^*\})=C_m(X')=\emptyset$.\\
	Case 2: If $\sum_{x\in X^{0\prime \prime}} s(a_x)< q_m$ but $Z^{0\prime \prime}=\emptyset$ then $Z^{0 \prime}\subseteq \{x^*\}$ while if $Z^{0 \prime}=\{x^*\}$ we have $C_m^{\prime}(X'\cup \{x^*\})=\{x^*\}$, leading to a contradiction. Hence $Z^{0 \prime}=Z^{0 \prime\prime}$ and $C_m(X'\cup \{x^*\})=C_m(X')$. 
	
	\noindent Otherwise, consider $a^{1\prime\prime}$ defined as $a\in A(Z^{0\prime\prime})$, such that $a \pi a'$ for all $a'\in  A(Z^{0\prime\prime})\setminus \{a\}$, and $x^{1\prime\prime}$ defined as $x\in Z_{a^{1\prime\prime}}^{0\prime\prime}$, such that $w_x<w_{x'}$ for all $x'\in Z^{0\prime\prime}_{a^{1\prime\prime}}\setminus \{x\}$.\\
	Suppose by contradiction that $x^{1\prime}\neq x^{1\prime \prime}$.
	Note that, by assumption $x^{1\prime}\neq x^*$ and by definition, we have $x^{1\prime}\in Z^{0 \prime}$.
	If $a_{x^{1\prime}}\neq a_{x^{1\prime\prime}}$, we reach a contradiction, as  $a_{x^{1\prime}}\in A(Z^{0\prime\prime})$ and therefore there exists a higher priority asylum seeker.
	Similarly, given $a_{x^{1\prime}}= a_{x^{1\prime\prime}}$ but $w_{x^{1\prime}}\neq w_{a^{1\prime\prime}}$, we reach a contradiction as $x^{1\prime}\in Z^{0\prime\prime}_{a^{1\prime\prime}}$ and the lowest wait time contract is uniquely defined. Finally, since $x^{1\prime}=x^{1\prime\prime }$ and $X^{0\prime}=X^{0\prime\prime}=\emptyset$ we have $X^{1\prime}=X^{0\prime}\cup \{x^{1\prime}\}= X^{0\prime\prime}\cup \{x^{1\prime\prime}\}=X^{1\prime\prime}$.

	\noindent \textbf{Induction step}: By the induction assumption, if the algorithm has not stopped at step $k-1$, we have $X^{k-1\prime}=X^{k-1 \prime \prime}$.
	We want to show that if the former algorithm stops at step $k$ we have $C_m(X'\cup \{x^*\})=C_m(X')$ and $X^{k\prime}=X^{k \prime \prime}$ otherwise.

	\noindent Suppose the former algorithm stops at step $k$.\\
	Case 1: If $\sum_{x\in X^{k-1\prime \prime}} s(a_x)\geq q_m$ then $\sum_{x\in X^{k-1 \prime}} s(a_x)\geq q_m$ as $X^{k-1\prime \prime}=X^{k-1\prime}$ and therefore $C_m(X'\cup \{x^*\})=C_m(X')=X^{k\prime\prime}$.\\
	Case 2: If $\sum_{x\in X^{k-1\prime \prime}} s(a_x)< q_m$ but $Z^{k\prime \prime}=\emptyset$ then $Z^{k \prime}=\subseteq \{x^*\}$. If $Z^{k \prime}=\{x^*\}$  we have $C_m^{\prime}(X'\cup \{x^*\})=\{x^*\}$, leading to a contradiction. Hence $Z^{k \prime}=Z^{k \prime\prime}$ and $C_m(X'\cup \{x^*\})=C_m(X')$. 
	
	\noindent Otherwise, consider $a^{k\prime\prime}$ defined as $a\in A(Z^{k\prime\prime})$, such that $a \pi a'$ for all $a'\in  A(Z^{k\prime\prime})\setminus \{a\}$, and $x^{k\prime\prime}$ defined as $x\in Z_{a^{k\prime\prime}}^{k\prime\prime}$, such that $w_x<w_{x'}$ for all $x'\in Z^{k\prime\prime}_{a^{k\prime\prime}}\setminus \{x\}$.\\
	Suppose by contradiction that $x^{k\prime}\neq x^{k\prime \prime}$.
	Note that by assumption $x^{k\prime}\neq x^*$ and moreover $Z^{k \prime}\setminus \{x^*\}=Z^{k \prime\prime}$.
	If $a_{x^{k\prime}}\neq a_{x^{k\prime\prime}}$, we reach a contradiction, as  $a_{x^{k+1\prime}}\in A(Z^{k\prime\prime})$ and the highest priority asylum seeker is uniquely defined.
	Similarly, given $a_{x^{k\prime}}= a_{x^{k\prime\prime}}$ but $w_{x^{k\prime}}\neq w_{a^{k\prime\prime}}$, we reach a contradiction, as $x^{k\prime}\in Z^{k\prime\prime}_{a^{k\prime\prime}}$ and the lowest wait time contract is uniquely defined. Finally, since $x^{k\prime}=x^{k\prime\prime }$ and by the induction assumption $X^{k\prime}=X^{k\prime\prime}$, we have $X^{k\prime}=X^{k-1\prime}\cup \{x^{k\prime}\}= X^{k-1\prime\prime}\cup \{x^{k\prime\prime}\}=X^{k\prime\prime}$. 
	
\end{proof}

\subsection*{\Cref{completion}(i)}		
\begin{proof}
	We have to show that for any $X'\subseteq X$, whenever there do not exist two contracts $x,x'\in C^{\prime}_{m}(X')$ specifying the same asylum seeker $a(x)=a(x')$ we must have $C^{\prime}_{m}(X')=C_{m}(X')$.
	The proof is by induction on contracts accepted during the steps of the algorithms describing the choice functions. 
	
	\noindent \textbf{Base Step:} If the algorithm describing $C_m$ stops at the end of step $1$ then $C^{\prime}_{m}(X')=C_{m}(X')$ and  $X^{1\prime}=X^{1}$ otherwise.  
	
	\noindent Suppose the algorithm stops at step $1$. \\
	By the definition of step $0$ we have $X^{0\prime}=X^{0}=\emptyset$ and therefore $Z^{0\prime}=\{x\in X'_m:r^{w_x}_m>|X^{0}_{w_x}|\text{ and } x\not\in X^{0})\}=Z^{0}=\{x\in X'_m:r^{w_x}_m>|X^{0}_{w_x}|\text{ and } a_x\not\in A(X^{0})\}$.
	Hence, if $\sum_{x\in X^{0}} s(a_x)\geq q_m$, then $\sum_{x\in X^{0\prime}} s(a_x)\geq q_m$ and both algorithms end. 
	Similarly, if $Z^{0}=\emptyset$ then $Z^{0\prime}=\emptyset$ and both algorithms end.
	In both instances  $C^{\prime}_{m}(X')=X^{0\prime}=X^0=C_{m}(X')$.
	
	\noindent Suppose the algorithm does not stops at step $1$. \\
	Under the original choice rule the highest priority asylum seeker in $a\in A(Z^0)$ is chosen, which is identical to the highest priority asylum seeker under $A(Z^{0\prime})=A(Z^{0})$, and hence $a^{1\prime}=a^{1}$. Moreover, the original choice rule selects the lowest wait time contact in $Z^{0}$, which is identical to the lowest wait time contract in $Z^{0\prime}=Z^{0}$, and hence $x^{1\prime}=x^1$.
	Trivially, we have $X^{1\prime}=X^{0\prime}\cup \{x^{1\prime}\}=X^{0}\cup \{x^{1}\}=X^{1}$.
	
	\noindent \textbf{Induction Step:} We assume that if the algorithm describing $C_m$ stops at the end of step $k-1$, then $C^{\prime}_{m}(X')=C_{m}(X')$ and  $X^{k-1\prime}=X^{k-1}$ otherwise.  \\
	We show that if the algorithm describing $C_m$ stops at the end of step $k$, then $C^{\prime}_{m}(X')=C_{m}(X')$ and $X^{k\prime}=X^{k}$ otherwise. 
	
	\noindent Suppose the algorithm stops at step $k$. \\
	By the induction assumption $X^{k-1\prime}=X^{k-1}$. Hence, if $\sum_{x\in X^{0}} s(a_x)\geq q_m$ then $\sum_{x\in X^{0\prime}} s(a_x)\geq q_m$ and both algorithms end, and by the induction assumption $C^{\prime}_{m}(X')=X^{k-1\prime}=X^{k-1}=C_{m}(X')$.
	
	\noindent By the induction assumption we have $X^{k-1\prime}=X^{k-1}$, as well as by definition $Z^{k-1\prime}=\{x\in X'_m:r^{w_x}_m>|X^{k-1}_{w_x}|\text{ and } x\not\in X^{k-1})\} \supseteq Z^{k-1}=\{x\in X'_m:r^{w_x}_m>|X^{k-1}_{w_x}|\text{ and } a_x\not\in A(X^{k-1})\}$. 
	Suppose there exist some $x\in Z^{k-1\prime}\setminus Z^{k-1}$. In this case $x$ belongs to an asylum seeker already being assigned a contract $a_x\in A(X^{k-1})$ with higher priority than any asylum seeker $a\in A(Z^{k-1})$. In other words, $x^{k\prime}\in Z^{k-1\prime}\setminus Z^{k-1}$ and at the same time we reach a contradiction with the initial assumption that there do not exist two accepted contracts specifying the same contracts, that is, there exists $x\in X^{k-1}$ with $a_{x}=a_{x^{k\prime}}$. We get $Z^{k-1\prime}=Z^{k-1}$.
	
	\noindent Suppose $Z^{k-1}=\emptyset$. Then $Z^{k-1\prime}=Z^{k-1}$ implies that $C^{\prime}_{m}(X')=X^{k-1\prime}=X^{k-1}=C_{m}(X')$.

	\noindent Suppose the algorithm does not stops at step $k$. \\
	Under the original choice rule the highest priority asylum seeker in $a\in A(Z^k)$ is chosen, which is identical to the highest priority asylum seeker under $A(Z^{k\prime})=A(Z^{k})$, and hence $a^{k\prime}=a^{k}$. Moreover, the original choice rule selects the lowest wait time contact in $Z^{k}$, which is identical to the lowest wait time contract in $Z^{k\prime}=Z^{k}$, and hence $x^{k\prime}=x^k$.
	Trivially, we have $X^{k\prime}=X^{k-1\prime}\cup \{x^{k\prime}\}=X^{k-1}\cup \{x^{k}\}=X^{k}$.
	
\end{proof}

\subsection*{\Cref{huso}}		
\Cref{huso} is helpful for analyzing the completion $C^{\prime}_{m}$. \Cref{huso} compares 
the contracts accepted under $C_m'(X)$ and  $C_m'(X\cup \{x^*\})$ given that $x^*\in C_m(X'\cup \{x^*\})$ and under the assumption that the burden-sharing stopping condition is not binding, that is, assuming $\sum_{x\in X'\cup \{x^*\}} s(a_x)\geq q_m$.
In essence, it states that there exists at most one previously accepted contract $x\in C_m'(X)$ that gets rejected $x\not \in C_m'(X\cup \{x^*\})$. 

\begin{lem}
	\label{huso}
	Assume  $\sum_{x\in X'\cup \{x^*\}} s(a_x)\geq q_m$.
	Consider $C_m'(X')$ and $C_m(X'\cup \{x^*\})$ for some $x^*\in X\setminus X'$ with $x^*\in C_m(X'\cup \{x^*\})$. 
	Let $\{x^{1\prime\prime}, \dotso, x^{K\prime\prime}\}=C_m'(X')$ and 
	$\{x^{1\prime}, \dotso, x^{l\prime}=x^*, \dotso, x^{K\prime}\}\subseteq C'_m(X\cup \{x^*\})$.
	Let step $\mathcal{l}_1$ be the step at which $x^*$ gets accepted. 
	If $|C_m'(X')_{w_{x^*}}|= r^{w_{x^*}}$, let step $\mathcal{l}_2$ be the step at which the last contract with wait time $w_{x^*}$ gets accepted ($\mathcal{l}_2\geq \mathcal{l}_1$). 
	
	\noindent Consider a contract $x^{j\prime}\neq x^*$ 
	\begin{itemize}
		\item[i)]   if $j<\mathcal{l}_1$ then $x^{j\prime}=x^{j\prime\prime}$,
		\item[ii)]   if $\mathcal{l}_2>j>\mathcal{l}_1$ then $x^{j\prime}=x^{j-1\prime \prime}$, and 
		\item[iii)]   if $j>\mathcal{l}_2$ then $x^{j\prime}=x^{j\prime\prime}$.
	\end{itemize}  
\end{lem}

\begin{proof}
		
	\noindent Consider $X'$ and $X'\cup \{x^*\}$ for some $x^*\in X\setminus X'$. 
	We refer to the relevant sets during each step of the algorithm for $C'_m$ as $X^{k\prime\prime}$, $Z^{k\prime\prime}$, and so on under the former ($C'_m(X')$) and $X^{k\prime}$, $Z^{k\prime}$, and so on under the latter ($C'_m(X'\cup \{x^*\})$). 
	
	\noindent For now, consider both algorithms but ignore the stopping point of either algorithm due to $\sum_{x\in X^{k-1\prime}} s(a_x)\geq q_m$ and $\sum_{x\in X^{k-1\prime\prime}} s(a_x)\geq q_m$, respectively. In other words, assume that $\sum_{x\in X^{\prime}\cup \{x^*\}} s(a_x)\geq q_m$.\\
	Moreover, assume that in step $l$ the latter accepts $x^{l\prime}=x^{*}$ and the former  $x^{l\prime\prime}$ and denote $w_{x^*}$ by $w^*$. Also, set the contracts accepted under the former as $\{x^{1\prime\prime}, \dotso, x^{l\prime \prime}, \dotso ,x^{l+k\prime \prime}, \dotso, x^{K\prime\prime}\}$.

	\noindent \textbf{Base Step.}  Consider step $l+1$.\\
	\noindent Case 1: If $|X^{l\prime \prime}_{w^*}|=|X^{l \prime}_{w^*}|=r^{w^{*}}_m$ and $|X^{l\prime \prime}_w|=|X^{l \prime}_w|$ for all $w\in W$, then $x^{l+1\prime}=x^{l+1\prime \prime}, \dotso, x^{K\prime}=x^{K\prime \prime}$.
	
	\noindent Case 2: If $|X^{l\prime \prime}_{w^*}|=|X^{l \prime}_{w^*}|<r^{w^{*}}_m$ and $|X^{l\prime \prime}_w|=|X^{l \prime}_w|$ for all $w\in W$, then $x^{l+1\prime}=x^{l\prime\prime}$ with $w^*=w_{x^{l\prime\prime}}$.
	
	
	\noindent Case 3: If $|X^{l\prime\prime}_{w_{x^{l\prime \prime}}}|=|X^{l\prime}_{w_{x^{l\prime \prime}}}|+1$, $|X^{l\prime}_{w^*}|=|X^{l\prime\prime}_{w^*}|+1$, and $|X^{l\prime \prime}_w|=|X^{l \prime}_w|$ for all $w\in W\setminus\{w^*,w_{x^{l\prime \prime}} \}$, then $x^{l+1\prime}=x^{l\prime\prime}$ with $w^*\neq w_{x^{l\prime\prime}}$.
	
	\noindent Up until step $l$ both algorithms accept identical contracts during each step, so $X^{l-1\prime}=X^{l-1 \prime \prime}$. Note that $Z^{k\prime}=\{x\in X_m':r^{x_w}_m>|X^{l\prime}_w| \text{ and } x\not\in X^{l\prime} \}$ as $X^{l\prime}=X^{l-1\prime}\cup \{x^*\}$ and $Z^{k\prime\prime}=\{x\in X_m' :r^{x_w}_m>|X^{l\prime\prime}_w| \text{ and } x\not\in X^{l\prime\prime} \}$ with $X^{l\prime\prime}=X^{l-1\prime}\cup \{x^{l\prime\prime}\}$.
	It follows that we can consider the following three relevant cases, covering every possible outcome.
	
	\noindent Case 1: $w^*=w_{x^{l\prime\prime}}$ and $|X^{l-1\prime\prime}_{w^*}|+1=|X^{l-1\prime}_{w^*}|+1=r^{w^*}_m$.
	
	\noindent We have $|X^{l\prime \prime}_{w^*}|=|X^{l \prime}_{w^*}|=r^{w^{*}}_m$, $|X^{l\prime \prime}_w|=|X^{l \prime}_w|$ for all $w\in W$, and $X^{l\prime\prime}\setminus\{x^{l\prime\prime}\}=X^{l\prime}\setminus\{x^{*}\}$. Hence,
	\begin{align*}
	Z^{l\prime} &=\{x\in X'_m: r^{x_w}_m>|X^{l\prime}_w|\text{ and } x\not\in X^{l\prime} \}\\
	&=\{x\in X'_m :r^{x_w}_m>|X^{l\prime\prime}_w|\text{ and } x\not\in X^{l\prime\prime} \}\\
	&=Z^{l\prime\prime}.
	\end{align*}
	It follows that every remaining step of both algorithms is identical, that is, $x^{l+1\prime}=x^{l+1\prime \prime}, \dotso, x^{K\prime}=x^{K\prime \prime}$. That is of course, everything is identical, except that $\sum_{x\in X^{l\prime}}s(a_x)$ might differ from $\sum_{x\in X^{l\prime\prime}}s(a_x)$, which is irrelevant, since by assumption $\sum_{x\in X^{\prime}} s(a_x)\geq q_m$.

	\noindent Case 2: $w^*=w^{l\prime\prime}$ and $|X^{l-1\prime\prime}_{w^*}|+1=|X^{l-1\prime}_{w^*}|<r^{w^{*}}_m$.

	\noindent We have $|X^{l\prime \prime}_{w^*}|=|X^{l \prime}_{w^*}|<r^{w^{*}}_m$, $|X^{l\prime \prime}_w|=|X^{l \prime}_w|$ for all $w\in W$, and $X^{l\prime\prime}\setminus\{x^{l\prime\prime}\}=X^{l\prime}\setminus\{x^{*}\}$. Hence,
	\begin{align*}
	Z^{l\prime} &=\{x\in X'_m: r^{x_w}_m>|X^{l\prime}_w|\text{ and } x\not\in X^{l\prime} \}\\
	&=\{x\in X'_m :r^{x_w}_m>|X^{l-1\prime\prime}_w|\text{ and } x\not\in X^{l-1\prime\prime} \}\\
	&=Z^{l-1\prime\prime}.
	\end{align*}
	Since $x^{l\prime \prime}$ was chosen due to having the highest priority and lowest wait time in $Z^{l-1\prime\prime}$, we have $x^{l+1\prime}=x^{l\prime \prime}$.

	\noindent Case 3: $w^*\neq w^{l\prime\prime}$.

	\noindent We have $|X^{l\prime\prime}_{w_{x^{l\prime \prime}}}|=|X^{l\prime}_{w_{x^{l\prime \prime}}}|+1$, $|X^{l\prime}_{w^*}|=|X^{l\prime\prime}_{w^*}|+1$,  $|X^{l\prime \prime}_w|=|X^{l \prime}_w|$ for all $w\in W\setminus\{w^*,w_{x^{l\prime \prime}}\}$, and $X^{l\prime\prime}\setminus\{x^{l\prime\prime}\}=X^{l\prime}\setminus\{x^{*}\}$. Hence,
	\begin{align*}
	Z^{k\prime}&=\{x\in X_m':r^{x_w}_m>|X^{l\prime}_w| \text{ and } x\not\in X^{l\prime} \}\\
	&\subseteq \{x\in X'_m:r^{x_w}_m>|X^{l-1\prime\prime}_w| \text{ and } x\not\in X^{l-1\prime}\}\\
	&=Z^{l-1\prime\prime}
	\end{align*}
	Since $x^{l\prime \prime}$ was chosen due to having the highest priority and lowest wait time in $Z^{l-1\prime\prime}$, we have $x^{l+1\prime}=x^{l\prime \prime}$ as $x^{l\prime \prime}\in Z^{l\prime}\subseteq Z^{l-1\prime\prime}$.

	\noindent \textbf{Induction Step.}\\
	Induction assumption: Assume there are cases 1, 2, and 3 at step $l+k$. 
	
	\noindent Case 1: If $|X^{l+k-1\prime \prime}_{w^*}|=|X^{l+k-1 \prime}_{w^*}|=r^{w^{*}}_m$ and $|X^{l+k-1\prime \prime}_w|=|X^{l+k-1 \prime}_w|$ for all $w\in W$ then  $x^{l+k\prime}=x^{l+k\prime \prime}, \dotso, x^{K\prime}=x^{K\prime \prime}$.
	
	\noindent Case 2: If $|X^{l+k-1\prime \prime}_{w^*}|=|X^{l+k-1 \prime}_{w^*}|<r^{w^{*}}_m$ and $|X^{l+k-1\prime \prime}_w|=|X^{l+k-1 \prime}_w|$ for all $w\in W$ then $x^{l+k\prime}=x^{l+k-1\prime\prime}$, with $w^*=w_{x^{l+k-1\prime\prime}}$.
	
	\noindent Case 3: If $|X^{l+k-1\prime\prime}_{w_{x^{l+k-1\prime \prime}}}|=|X^{l+1\prime}_{w_{x^{l+k-1\prime \prime}}}|+1$, $|X^{l+k-1\prime}_{w^*}|=|X^{l+k-1\prime\prime}_{w^*}|+1$ and $|X^{l+k-1\prime \prime}_w|=|X^{l+k-1 \prime}_w|$ for all $w\in W\setminus\{w^*,w_{x^{l\prime \prime}} \}$ then $x^{l+k\prime}=x^{l+k-1\prime\prime}$, with $w^*\neq w_{x^{l+k-1\prime\prime}}$.
	
	\noindent There are the same cases 1, 2, and 3 at step $l+k+1$.

	\noindent 	\textbf{Case 1} holds at step $l+k$.\\
Note that at step $l+k$ if we are in case $1$ then trivially case $1$ holds for step $l+k+1$.

	\noindent 	\textbf{Case 2} holds at step $l+k$.\\
	By the induction assumption, $|X^{l+k-1\prime \prime}_{w^*}|=|X^{l+k-1 \prime}_{w^*}|<r^{w^{*}}_m$ ,$|X^{l+k-1\prime \prime}_w|=|X^{l+k-1 \prime}_w|$ for all $w\in W$, and we have $x^{l+k\prime}=x^{l+k-1\prime\prime}$, with $w^*=w_{x^{l+k-1\prime\prime}}$.
	
	\noindent Case 1: We are in case 1 if $w_{x^{l+k\prime\prime}}=w^*$ and $|X^{l+k-1\prime \prime}_{w^*}|+1=|X^{l+k-1\prime}_{w^*}|+1=r^{w^*}_m$.

	\noindent We have $|X^{l+k\prime \prime}_{w^*}|=|X^{l+k \prime}_{w^*}|=r^{w^{*}}_m$, $|X^{l+k\prime \prime}_w|=|X^{l+k \prime}_w|$ for all $w\in W$, and $X^{l+k\prime\prime}\setminus \{x^{l+k\prime\prime}\}=X^{l+k\prime}\setminus \{x^{*}\}$. Hence,
	\begin{align*}
	Z^{l+k\prime} &=\{x\in X'_m: r^{x_w}_m>|X^{l+k\prime}_w|\text{ and } x\not\in X^{l+k\prime} \}\\
	&=\{x\in X'_m:r^{x_w}_m>|X^{l+k\prime\prime}_w|\text{ and } x\not\in X^{l+k\prime \prime}\}\\
	&=Z^{l+k\prime\prime}.
	\end{align*}
	Therefore, all remaining accepted contracts are identical, that is $x^{l+k+1\prime}=x^{l+k+1\prime \prime}, \dotso, x^{K\prime}=x^{K\prime \prime}$.

	\noindent Case 2:
	We are in case 2 if $w_{x^{l+k\prime\prime}}=w^*$ and $|X^{l+k-1\prime \prime}_{w^*}|+1=|X^{l+k-1\prime}_{w^*}|+1=r^{w^*}_m$.
	
	\noindent We have $|X^{l+k\prime \prime}_{w^*}|=|X^{l+k \prime}_{w^*}|<r^{w^{*}}_m$, $|X^{l+k\prime \prime}_w|=|X^{l+k \prime}_w|$ for all $w\in W$, and $X^{l+k\prime\prime}\setminus \{x^{l+k\prime\prime}\}=X^{l+k\prime}\setminus \{x^{*}\}$. Hence,
	\begin{align*}
	Z^{l+k\prime} &=\{x\in X'_m: r^{x_w}_m>|X^{l+k\prime}_w|\text{ and } x\not\in X^{l+k\prime} \}\\
	&=\{x\in X'_m:r^{x_w}_m>|X^{l+k-1\prime\prime}_w|\text{ and } x\not\in X^{l+k-1\prime \prime}\}\\
	&=Z^{l+k-1\prime\prime}.
	\end{align*}
	Since $x^{l+k\prime \prime}$ was chosen due to having the highest priority and lowest wait time in $Z^{l-1\prime\prime}$, we have $x^{l+k+1\prime}=x^{l+k\prime \prime }$.

	\noindent Case 3:
	We are in case 3 if $w_{x^{l+k\prime\prime}}\neq w^*$.
	
	\noindent We have $|X^{l+k\prime\prime}_{w_{x^{l+k\prime \prime}}}|=|X^{l+k\prime}_{w_{x^{l+k\prime \prime}}}|+1$, $|X^{l+k\prime}_{w^*}|=|X^{l+k\prime\prime}_{w^*}|+1$, $|X^{l+k\prime \prime}_w|=|X^{l+k \prime}_w|$ for all $w\in W\setminus\{w^*,w_{x^{l+k\prime \prime}} \}$, and $X^{l+k\prime\prime}\setminus \{x^{l+k\prime\prime}\}=X^{l+k\prime}\setminus \{x^{*}\}$. Hence,
	
	\begin{align*}
	Z^{l+k\prime} &=\{x\in X'_m: r^{x_w}_m>|X^{l+k\prime}_w|\text{ and } x\not\in X^{l+k\prime} \}\\
	&\subseteq \{x\in X'_m:r^{x_w}_m>|X^{l+k-1\prime\prime}_w|\text{ and } x\not\in X^{l+k-1\prime \prime}\}\\
	&=Z^{l+k-1\prime\prime}.
	\end{align*}
	Since $x^{l+k\prime \prime}$ was chosen due to having the highest priority and lowest wait time in $Z^{l+k-1\prime\prime}$, we have $x^{l+k+1\prime}=x^{l+k\prime \prime}$ as $x^{l+k\prime \prime}\in Z^{l+k\prime}\subseteq Z^{l+k-1\prime\prime}$.

	\noindent 	\textbf{Case 3} holds at step $l+k$.\\
	By the induction assumption, $|X^{l+k-1\prime\prime}_{w_{x^{l+k-1\prime \prime}}}|=|X^{l+k-1\prime}_{w_{x^{l+k-1\prime \prime}}}|+1$, $|X^{l+k-1\prime}_{w^*}|=|X^{l+k-1\prime\prime}_{w^*}|+1$, $|X^{l+k-1\prime \prime}_w|=|X^{l+k-1 \prime}_w|$ for all $w\in W\setminus\{w^*,w_{x^{l\prime \prime}} \}$ and we have $x^{l+k\prime}=x^{l+k-1\prime\prime}$, with $w^*\neq w_{x^{l+k-1\prime\prime}}$.

	\noindent Case 1:
	We are in case 1 if $w_{x^{l+k\prime\prime}}=w^*$ and $|X^{l+k-1\prime \prime}_{w^*}|+1=r^{w^*}_m$.
	
	\noindent We have $|X^{l+k\prime \prime}_{w^*}|=|X^{l+k \prime}_{w^*}|=r^{w^{*}}_m$, $|X^{l+k\prime \prime}_w|=|X^{l+k \prime}_w|$ for all $w\in W$and $X^{l+k\prime\prime}\setminus \{x^{l+k\prime\prime}\}=X^{l+k\prime}\setminus \{x^{*}\}$. 
	
	\noindent The remaining argument for $x^{l+k+1\prime}=x^{l+k+1\prime \prime}, \dotso, x^{K\prime}=x^{K\prime \prime}$ is identical to the previous argument for case $1$ as $Z^{l+k\prime}=Z^{l+k\prime\prime}$.

	\noindent Case 2:
	We are in case 2 if $w_{x^{l+k\prime\prime}}=w^*$ and $|X^{l+k-1\prime \prime}_{w^*}|+1<r^{w^*}_m$.
	
	\noindent We have $|X^{l+k\prime \prime}_{w^*}|=|X^{l+k \prime}_{w^*}|<r^{w^{*}}_m$ and $|X^{l+k\prime \prime}_w|=|X^{l+k \prime}_w|$ for all $w\in W$, and $X^{l+k\prime\prime}\setminus \{x^{l+k\prime\prime}\}=X^{l+k\prime}\setminus \{x^{*}\}$. 
	
	\noindent The remaining argument for $x^{l+k+1\prime}=x^{l+k\prime \prime}$ is identical to the previous argument for case $2$, as $Z^{l+k\prime}=Z^{l+k-1\prime\prime}$.

	\noindent Case 3:
	We are in case 3 if $w_{x^{l+k\prime\prime}}\neq w^*$.
	
	\noindent  We have $|X^{l+k\prime\prime}_{w_{x^{l+k\prime\prime}}}|=|X^{l+k\prime}_{w_{x^{l+k\prime\prime}}}|+1$, $|X^{l+k\prime\prime}_{w^*}|+1=|X^{l+k\prime}_{w^*}|$, $|X^{l+k\prime \prime}_w|=|X^{l+k\prime}_w|$ for all $w\in W \setminus \{w^*,w_{x^{l+k\prime\prime}}\}$, and $X^{l+k\prime\prime}\setminus \{x^{l+k\prime\prime}\}=X^{l+k\prime}\setminus \{x^{*}\}$. 
	 
	\noindent The remaining argument for $x^{l+k+1\prime}=x^{l+k\prime \prime}$ is identical to the previous argument for case $3$, as $x^{l+k\prime \prime}\in Z^{l+k\prime}\subseteq Z^{l+k-1\prime\prime}$.
	
\end{proof}

	\subsection*{\Cref{completion}(ii) }
	
	\begin{proof}
		
		\noindent This proof is almost identical to the proof for \Cref{IRC}.
		 Consider $X'$ and $X'\cup \{x^*\}$ for some $x^*\in X\setminus X'$. 
		We refer to the relevant sets during each step of the algorithm for $C'_m$ as $X^{k\prime\prime}$, $Z^{k\prime\prime}$, and so on under the former ($C'_m(X')$) and $X^{k\prime}$, $Z^{k\prime}$, and so on under the latter ($C'_m(X'\cup \{x^*\})$). 
		
		\noindent Given that $x^*\not \in C_m'(X'\cup \{x^*\})$, we want to show that $C_m^{\prime}(X'\cup \{x^*\})=C^{\prime}_m(X')$. We proceed by induction.
		
		\noindent \textbf{Base step}: We assume that $x^*\not \in C_m^{\prime}(X'\cup \{x^*\})$. If the former algorithm ($C'_m(X')$) stops at step $1$, we have $C_m^{\prime}(X'\cup \{x^*\})=C_m^{\prime}(X')$ and $X^{1\prime}=X^{1 \prime \prime}$ otherwise.
		
		
		\noindent Suppose the former algorithm stops at step $1$.\\
		Case 1: If $\sum_{x\in X^{0\prime \prime}} s(a_x)\geq q_m$, then $\sum_{x\in X^{0 \prime}} s(a_x)\geq q_m$ as $X^{0\prime \prime}=X^{0\prime}=\emptyset$ and therefore $C'_m(X'\cup \{x^*\})=C'_m(X')=\emptyset$.\\
		Case 2: If $\sum_{x\in X^{0\prime \prime}} s(a_x)< q_m$ but $Z^{0\prime \prime}=\emptyset$, then $Z^{0 \prime}\subseteq \{x^*\}$ while if $Z^{0 \prime}=\{x^*\}$ we have $C_m^{\prime}(X'\cup \{x^*\})=\{x^*\}$, which leads to a contradiction. Hence $Z^{0 \prime}=Z^{0 \prime\prime}$ and $C'_m(X'\cup \{x^*\})=C'_m(X')$. 
		
		\noindent Otherwise, consider $a^{1\prime\prime}$ defined as $a\in A(Z^{0\prime\prime})$, such that $a \pi a'$ for all $a'\in  A(Z^{0\prime\prime})\setminus \{a\}$, and $x^{1\prime\prime}$ defined as $x\in Z_{a^{1\prime\prime}}^{0\prime\prime}$, such that $w_x<w_{x'}$ for all $x'\in Z^{0\prime\prime}_{a^{1\prime\prime}}\setminus \{x\}$.\\
		Suppose by contradiction that $x^{1\prime}\neq x^{1\prime \prime}$.
		Note that, by assumption $x^{1\prime}\neq x^*$ and by definition, we have $x^{1\prime}\in Z^{0 \prime}$.
		If $a_{x^{1\prime}}\neq a_{x^{1\prime\prime}}$ we reach a contradiction, as  $a_{x^{1\prime}}\in A(Z^{0\prime\prime})$ and therefore there exists a higher priority asylum seeker.
		Similarly, given $a_{x^{1\prime}}= a_{x^{1\prime\prime}}$ but $w_{x^{1\prime}}\neq w_{a^{1\prime\prime}}$, we reach a contradiction, as $x^{1\prime}\in Z^{0\prime\prime}_{a^{1\prime\prime}}$ and the lowest wait time contract is uniquely defined. Finally, since $x^{1\prime}=x^{1\prime\prime }$ and $X^{0\prime}=X^{0\prime\prime}=\emptyset$, we have $X^{1\prime}=X^{0\prime}\cup \{x^{1\prime}\}= X^{0\prime\prime}\cup \{x^{1\prime\prime}\}=X^{1\prime\prime}$.

		\noindent \textbf{Induction step}: By the induction assumption, if the algorithm has not stopped at step $k-1$, we have $X^{k-1\prime}=X^{k-1 \prime \prime}$.
		We want to show that if the former algorithm stops at step $k$, we will have $C'_m(X'\cup \{x^*\})=C'_m(X')$ and $X^{k\prime}=X^{k \prime \prime}$ otherwise.

		\noindent Suppose the former algorithm stops at step $k$.\\
		Case 1: If $\sum_{x\in X^{k-1\prime \prime}} s(a_x)\geq q_m$, then $\sum_{x\in X^{k-1 \prime}} s(a_x)\geq q_m$ as $X^{k-1\prime \prime}=X^{k-1\prime}$ and therefore $C'_m(X'\cup \{x^*\})=C'_m(X')=X^{k\prime\prime}$.\\
		Case 2: If $\sum_{x\in X^{k-1\prime \prime}} s(a_x)< q_m$ but $Z^{k\prime \prime}=\emptyset$, then $Z^{k \prime}=\subseteq \{x^*\}$. If $Z^{k \prime}=\{x^*\}$, we have $C_m^{\prime}(X'\cup \{x^*\})=\{x^*\}$, which leads to a contradiction. Hence $Z^{k \prime}=Z^{k \prime\prime}$ and $C'_m(X'\cup \{x^*\})=C'_m(X')$. 
		
		\noindent Otherwise, consider $a^{k\prime\prime}$ defined as $a\in A(Z^{k\prime\prime})$, such that $a \pi a'$ for all $a'\in  A(Z^{k\prime\prime})\setminus \{a\}$, and $x^{k\prime\prime}$ defined as $x\in Z_{a^{k\prime\prime}}^{k\prime\prime}$ such that $w_x<w_{x'}$ for all $x'\in Z^{k\prime\prime}_{a^{k\prime\prime}}\setminus \{x\}$.\\
		Suppose by contradiction that $x^{k\prime}\neq x^{k\prime \prime}$.
		Note that, by assumption $x^{k\prime}\neq x^*$ and moreover $Z^{k \prime}\setminus \{x^*\}=Z^{k \prime\prime}$.
		If $a_{x^{k\prime}}\neq a_{x^{k\prime\prime}}$, we reach a contradiction, as  $a_{x^{k+1\prime}}\in A(Z^{k\prime\prime})$ and the highest priority asylum seeker is uniquely defined.
		Similarly, given $a_{x^{k\prime}}= a_{x^{k\prime\prime}}$ but $w_{x^{k\prime}}\neq w_{a^{k\prime\prime}}$, we reach a contradiction as $x^{k\prime}\in Z^{k\prime\prime}_{a^{k\prime\prime}}$ and the lowest wait time contract is uniquely defined. Finally, since $x^{k\prime}=x^{k\prime\prime }$ and by the induction assumption $X^{k\prime}=X^{k\prime\prime}$, we have $X^{k\prime}=X^{k-1\prime}\cup \{x^{k\prime}\}= X^{k-1\prime\prime}\cup \{x^{k\prime\prime}\}=X^{k\prime\prime}$. 
		
		\noindent By \Cref{completion}(i), $C^{\prime}_{m}$ is a completion of $C_{m}(X')$, and therefore there exists a completion satisfying  irrelevance of rejected contracts. 
	\end{proof}

\subsection*{\Cref{completion}(iv)}		
\begin{proof}
Consider $X'$ and $X'\cup \{x^*\}$ for some $x^*\in X\setminus X'$. We refer to the relevant sets during each step of the algorithm for $C'_m$ as $X^{k\prime\prime}$, $Z^{k\prime\prime}$, and so on under the former ($C'_m(X')$) and $X^{k\prime}$, $Z^{k\prime}$, and so on under the latter ($C'_m(X'\cup \{x^*\})$). 
	
	\noindent Case 1. Suppose that $x^*\not \in C_m'(X'\cup \{x^*\})$.\\ 
	By irrelevance of rejected contracts $C_m'(X'\cup \{x^*\})=C_m'(X' )$ and therefore $|C_m'(X'\cup \{x^*\})|=|C_m'(X' )|$.
	
	\noindent Case 2. Suppose that $x^*\in C_m'(X'\cup \{x^*\})$.
	
	\noindent Consider a step $j$ with $\mathcal{l}_2\geq j>\mathcal{l}_1$ where we have $X^{j-1\prime\prime}=\{x^{1\prime \prime}\dotso ,x^{\mathcal{l}_1\prime\prime}, \dotso, x^{j-1\prime\prime}\}$ and by \Cref{huso} we get that $X^{j-1\prime}=\{x^{1\prime \prime}\dotso ,x^{*},x^{\mathcal{l}_1\prime\prime}, \dotso, x^{j-2\prime\prime}\}$. 
	In other words, $X^{j-1\prime\prime}\setminus \{x^{j-1\prime \prime}\}=X^{j-1\prime}\setminus \{x^*\}$ and by small burden-size priority $s(a_{x^*})\leq s(a_{x^{j-1\prime \prime}})$ and therefore $\sum_{x\in X^{j-1\prime}} s(a_x)\leq \sum_{x\in X^{j-1\prime\prime}}s(a_x)$.\\
	 It follows that if another step is taken under $X'$, then another step is taken under $X'\cup \{x\}$.
	 
	\noindent Similarly, for any step $j$ with $j>\mathcal{l}_2$, we have $X^{j-1\prime\prime}=\{x^{1\prime \prime}\dotso ,x^{\mathcal{l}_1\prime\prime}, \dotso, x^{\mathcal{l}_2\prime\prime}, \dotso, x^{j-1\prime\prime} \}$ and by \Cref{huso} we get that $X^{j-1\prime}=\{x^{1\prime \prime},\dotso ,x^{*},x^{\mathcal{l}_1\prime\prime}, \dotso, x^{\mathcal{l}_2\prime}=x^{\mathcal{l}_2-1\prime\prime}, x^{\mathcal{l}_2+1\prime}=x^{\mathcal{l}_2+1\prime \prime}, \dotso, x^{j-1\prime}=x^{j-1\prime\prime} \}$.\\
	In other words, $X^{j-1\prime\prime}\setminus \{x^{\mathcal{l}_2\prime\prime}\}=X^{j-1\prime} \setminus \{x^*\}$ with $a_{x^*}\pi a_{x^{\mathcal{l}_2\prime\prime}}$ and by small burden-size priority $s(a_{x^*})\leq s(x^{\mathcal{l}_2\prime\prime})$ and therefore $\sum_{x\in X^{j-1\prime}} s(a_x)\leq \sum_{x\in X^{j-1\prime\prime}}s(a_x)$. \\
	 It follows that if another step is taken under $X'$, then another step is taken under $X'\cup \{x\}$.
	 
	 \noindent To sum up, under small burden-size, if another step is taken under the algorithm for $X'$, then another step is taken under the algorithm for $X'\cup \{x\}$. Since each step corresponds to an accepted contract, we have $|C^{\prime}_{m}(X'\cup \{x^* \})|\geq |C^{\prime}_{m}(X')|$.
		
		\noindent By \Cref{completion}(i), $C^{\prime}_{m}$ is a completion of $C_{m}(X')$, and therefore there exists a completion satisfying the law of aggregate demand.

\end{proof}

\subsection*{\Cref{completion}(iii)}		
\begin{proof}
	\noindent  Consider $X'$ and $X'\cup \{x^*\}$ for some $x^*\in X\setminus X'$. 
	We refer to the relevant sets during each step of the algorithm for $C'_m$ as $X^{k\prime\prime}$, $Z^{k\prime\prime}$, and so on under the former ($C'_m(X')$) and $X^{k\prime}$, $Z^{k\prime}$, and so on under the latter ($C'_m(X'\cup \{x^*\})$). 
	
	\noindent Case 1. Suppose that $x^*\not \in C_m'(X'\cup \{x^*\})$.\\ 
	By irrelevance of rejected contracts $C_m'(X'\cup \{x^*\})=C_m'(X' )$ and therefore if $x\in  C_m'(X'\cup \{x^*\})$ then $x\in  C_m'(X')$.
	
	\noindent Case 2. Suppose that $x^*\in C_m'(X'\cup \{x^*\})$.\\
	
	\noindent Consider a step $j$ with $\mathcal{l}_2\geq j>\mathcal{l}_1$ where we have $X^{j-1\prime\prime}=\{x^{1\prime \prime}\dotso ,x^{\mathcal{l}_1\prime\prime}, \dotso, x^{j-1\prime\prime}\}$ and by \Cref{huso} we get that $X^{j-1\prime}=\{x^{1\prime \prime}\dotso ,x^{*},x^{\mathcal{l}_1\prime\prime}, \dotso, x^{j-2\prime\prime}\}$. \\
	In other words, $X^{j-1\prime\prime}\setminus \{x^{j-1\prime \prime}\}=X^{j-1\prime}\setminus \{x^*\}$ and by large burden-size priority $s(a_{x^*})\geq s(a_{x^{j-1\prime \prime}})$ and therefore $\sum_{x\in X^{j-1\prime}} s(a_x)\geq \sum_{x\in X^{j-1\prime\prime}}s(a_x)$.\\
	It follows that if another step is taken under $X'\cup \{x\}$, then another step is taken under $X'$ --- large-size priority reverses the observation under \Cref{completion}(iv).
	
	\noindent Similarly, for any step $j$ with $j>\mathcal{l}_2$, we have $X^{j-1\prime\prime}=\{x^{1\prime \prime}\dotso ,x^{\mathcal{l}_1\prime\prime}, \dotso, x^{\mathcal{l}_2\prime\prime}, \dotso, x^{j-1\prime\prime} \}$ and by \Cref{huso} we get that $X^{j-1\prime}=\{x^{1\prime \prime},\dotso ,x^{*},x^{\mathcal{l}_1\prime\prime}, \dotso, x^{\mathcal{l}_2\prime}=x^{\mathcal{l}_2-1\prime\prime}, x^{\mathcal{l}_2+1\prime}=x^{\mathcal{l}_2+1\prime \prime}, \dotso, x^{j-1\prime}=x^{j-1\prime\prime} \}$.
	In other words, $X^{j-1\prime\prime}\setminus \{x^{\mathcal{l}_2\prime\prime}\}=X^{j-1\prime} \setminus \{x^*\}$ with $a_{x^*}\pi a_{x^{\mathcal{l}_2\prime\prime}}$ and by large burden-size priority $s(a_{x^*})\leq s(a_{x^{**}})$ and therefore $\sum_{x\in X^{j-1\prime}} s(a_x)\geq \sum_{x\in X^{j-1\prime\prime}}s(a_x)$. \\
	It follows that if another step is taken under $X'\cup \{x\}$, then another step is taken under $X'$.
	
	\noindent To sum up, under large burden-size, if another step is taken under the algorithm for $X'\cup \{x\}$, then another step is taken under the algorithm for $X'$. Therefore, if $x\in C^{\prime}_{m}(X'\cup \{x^*\}) \setminus \{x^*\}$ then $x\in C^{\prime}_{m}(X')$
	
	\noindent By \Cref{completion}(i), $C^{\prime}_{m}$ is a completion of $C_{m}(X')$, and therefore there exists a completion satisfying substitutability.

\end{proof}

\subsection*{\Cref{NOM for large burden size priority}}		

\begin{proof}
By \Cref{completion}, under large burden size priority the asylum-seeker-proposing cumulative offer mechanism leads a stable outcome as the completion $\hat{C}'_m$ satisfies substitutability and irrelevance of rejected contracts. 
\medskip

\noindent It remains to be shown that the asylum-seeker-proposing cumulative offer mechanism is non-obviously manipulable if priorities satisfy large burden size priority.  
We show that for a given true preference $P_a$, no asylum seeker can submit another preference $\hat{P}_a$ that either gives a strictly better best outcome or a strictly better worst outcome. \medskip

\noindent \textbf{Best possible outcome.} It is obvious that, under any asylum seeker $a$'s true preferences $P_a$ she can get her top choice for some $P'_{-a} \in \mathcal{P}_{-a}$. Thus, for any preference $\hat{P}_a \in \mathcal{P}_a$ we have $W_\varphi(a;P_a)  \;R_a\; W_\varphi(a;\hat{P}_a)$, contradicting that any profitable manipulation can achieve $W_\varphi(a;\hat{P}_a) \;P_a\; W_\varphi(a;P_a)$.
\medskip

\noindent \textbf{Worst possible outcome.} Let  $a$'s worst outcome be achieved under some preference profile $P^w_{-a} \in \mathcal{P}_{-a}$, i.e.,  $W_{\varphi}(a,P_a)=\varphi(s,P_a,P^w_{-a})_a$.
Now consider some profitable manipulation $\hat{P}_a$ and suppose it yields a strictly better worst outcome $W_{\varphi}(a,\hat{P}_a) \;P_a \; W_{\varphi}(a,P_a)$. 
We show that no such manipulation can exist. In particular, consider the following preference profile $\hat{P}_{-a} \in \mathcal{P}_{-a}$, where the other asylum seekers $a'\in A \setminus\{a\}$ submit the following preferences:
$$\hat{P}_{a'}:  \varphi(s,P_a,P^w_{-a})_{a'} - \varphi(s,P_a,P^w_{-a})_{a} - \ldots$$


\noindent Let every $a'\in A$ propose before $a$ and note that everyone's contract is tentatively accepted.
Moreover, note the following, given any $P'_a \in \mathcal{P}_a$ and in particular $\hat{P}_a$:
\begin{itemize}
    \item[i)] If $a$ proposes a contract $x$ that is strictly preferred to the worst possible outcome under truthful preference $\varphi(s,P_a,P^w_{-a})_a$; that contract must be rejected as otherwise we have a contradiction with stability --- as, given \Cref{completion}, under large burden-size priority, $\hat{C}'_m$ satisfies substitutability and irrelevance of rejected contracts . Given irrelevance of rejected contracts proposing such a contract does not change the outcome of the COM.
    \item[ii)] If $a$ proposes $\varphi(s,P_a,P^w_{-a})_a$ itself, the COM ends and $\hat{P}_a$ has not led to a strictly better outcome.
    \item[iii.a)] If $a$ proposes a strictly worse contract and the contract is rejected it does not affect the outcome of the COM by irrelevance of rejected contracts. 
     \item[iii.b)] If the contract is accepted, $a$ will be assigned that contract at the end of the COM. 
\begin{itemize}
    \item     If the acceptance of the contract leads to no rejection the COM trivially ends as every asylum seeker has a contract assigned.
    \item    If the  acceptance of $a$'s contracts leads to a rejection of an asylum seeker with the same burden size; that asylum seeker will apply to $\varphi(s,P_a,P^w_{-a})_a$, which will be accepted and leads to a stop of the COM --- therefore, $a$ ends up with a worse contract than $\varphi(s,P_a,P^w_{-a})_a$.
    \item  If the acceptance leads to the rejection of more than one asylum seeker, those asylum seekers must have a lower burden size than $a$, and therefore, by large burden size priority, a lower priority than $a$ or anyone else having a higher priority than $a$ at some or all member states. It follows that no matter what contracts the rejected asylum seekers propose, $a$ can never be rejected from her current contract --- therefore, $a$ ends up with a worse contract than $\varphi(s,P_a,P^w_{-a})_a$.
\end{itemize} 
\end{itemize}

\noindent Putting $i) -iii)$ together we find a contradiction with $W_\varphi(a;\hat{P}_a) \;P_a\; W_\varphi(a;P_a)$ as \\$W_\varphi(a;P_a) \;R_a\; W_\varphi(a;\hat{P}_a)$, concluding the proof.

\end{proof}

\end{document}

%% file: library2.bib

@article{roberts1985measurement,
  title={Measurement theory},
  author={Roberts, Fred S},
  year={1985}
}


@article{pentico2007assignment,
  title={Assignment problems: A golden anniversary survey},
  author={Pentico, David W},
  journal={European Journal of Operational Research},
  volume={176},
  number={2},
  pages={774--793},
  year={2007},
  publisher={Elsevier}
}


@article{HumanR,
  title={European Convention on Human Rights},
  author={{European Council of Human Rights}},
url = {https://www.echr.coe.int/Documents/Convention_ENG.pdf},
  year={2013}
}

@article{Schuck1997,
  title={Refugee Burden-Sharing: A Modest Proposal},
  author={Schuck, Peter H},
  journal={Yale Journal of International Law},
  volume={22},
  year={1997}
}

@Article{Morgana2014,
  author    = {Hagen, Martin},
  journal   = {Journal of Public Economics},
  title     = {Tradable immigration quotas revisited},
  year      = {2022},
  pages     = {104619},
  volume    = {208},
  publisher = {Elsevier},
}

@Article{fernandez2015tradable,
  author    = {Fern{\'a}ndez-Huertas Moraga, Jes{\'u}s and Rapoport, Hillel},
  journal   = {CESifo Economic Studies},
  title     = {Tradable refugee-admission quotas and EU asylum policy},
  year      = {2015},
  number    = {3-4},
  pages     = {638--672},
  volume    = {61},
  publisher = {Oxford University Press},
}

@Article{Jones2016,
  author    = {Jones, Will and Teytelboym, Alexander},
  journal   = {Forced Migration Review},
  title     = {Choices, preferences and priorities in a matching system for refugees},
  year      = {2016},
  number    = {51},
  publisher = {Refugee Studies Centre, University of Oxford},
}

@Article{Teytelboym,
  author    = {Jones, Will and Teytelboym, Alexander},
  journal   = {Journal on Migration and Human Security},
  title     = {Matching systems for refugees},
  year      = {2017},
  number    = {3},
  pages     = {667--681},
  volume    = {5},
  publisher = {SAGE Publications Sage CA: Los Angeles, CA},
}

@article{jones2017international,
  title={The international refugee match: A system that respects refugees’ preferences and the priorities of states},
  author={Jones, Will and Teytelboym, Alexander},
  journal={Refugee Survey Quarterly},
  volume={36},
  number={2},
  pages={84--109},
  year={2017},
  publisher={Oxford University Press}
}

@Article{Delacretaz2016,
  author    = {Delacr{\'e}taz, David and Kominers, Scott Duke and Teytelboym, Alexander},
  journal   = {American Economic Review},
  title     = {Matching mechanisms for refugee resettlement},
  year      = {2023},
  number    = {10},
  pages     = {2689--2717},
  volume    = {113},
  publisher = {American Economic Association 2014 Broadway, Suite 305, Nashville, TN 37203},
}

@Article{andersson2018dynamic,
  author      = {Andersson, Tommy and Ehlers, Lars and Martinello, Alessandro},
  title       = {Dynamic refugee matching},
  year        = {2018},
  institution = {Working Paper},
}

@Article{trapp2018placement,
  author    = {Ahani, Narges and Andersson, Tommy and Martinello, Alessandro and Teytelboym, Alexander and Trapp, Andrew C},
  title     = {Placement optimization in refugee resettlement},
  year      = {2021},
  number    = {5},
  journal   = {Operations Research},
  pages     = {1468--1486},
  publisher = {INFORMS},
  volume    = {69},
}


@techreport{EUProp2016,
author = {{European Commission}},
title = {Proposal for 
establishing the criteria and mechanisms for determining the Member State responsible
for examining an application for international protection lodged in one of the Member
States by a third-country national or a stateless person },
url = {https://ec.europa.eu/transparency/regdoc/rep/1/2016/EN/1-2016-270-EN-F1-1.PDF},
year = {2016}
}

@article{beck2014application,
  title={The application of the EU Charter of Fundamental Rights to asylum procedure law, European Council on Refugees and Exiles},
  author={Beck, Gunnar and Mole, Nuale and Reneman, Marcelle},
  year={2014},
  publisher={Dutch Council of Refugees}
}

@Article{neville2016towards,
  author    = {Neville, Darren and Rigon, Amalia},
  title     = {Towards an EU humanitarian visa scheme?},
  year      = {2016},
  publisher = {EPRS: European Parliamentary Research Service},
}

@TechReport{EUREG2016,
  author      = {Maiani, Francesco},
  institution = {Parlement europ{\'e}en},
  title       = {The Reform of the Dublin III Regulation},
  year        = {2016},
}

@Misc{Seema2018,
  author  = {Jilani, Seema},
  title   = {What refugees face on the world’s deadliest migration route},
  year    = {2018},
  journal = {The New York Times Magazine},
}

{huffingtonpost2018,
author = {Gray, Jasmin},
booktitle = {Huffington Post},
title = {{'My Life Is Passing Me By': What It's Like To Wait Years For An Asylum Decision}},
url = {https://www.huffingtonpost.co.uk/entry/wait-asylum-decision_uk_5afacd27e4b09a94524c06e8?guccounter=1},
year = {2018}

@Misc{Financial2015,
  author  = {Wasik, Zosia and Foy, Henry},
  title   = {Poland favours Christian refugees from Syria},
  year    = {2015},
  journal = {Financial Times},
  number  = {8},
  pages   = {2015},
  volume  = {21},
}

@Misc{BBC2016,
  author  = {Margaroni, Maria},
  title   = {Greece’s Stranded Refugees Fear Being Forgotten},
  year    = {2016},
  journal = {BBC World Service},
  volume  = {7},
}

@Misc{Guardian2015,
  author  = {Harding, Luke},
  title   = {Refugee crisis: Germany reinstates controls at Austrian border},
  year    = {2015},
  journal = {The Guardian},
  volume  = {13},
}

@misc{Vicenews2015,
author = {Hayden, Sally},
booktitle = {ViceNews},
title = {{Germany Is Set to Accept Asylum Applications From all Syrians Who Arrive There}},
url = {https://news.vice.com/article/germany-is-set-to-accept-asylum-applications-from-all-syrians-who-arrive-there},
year = {2015}
}

@misc{QuotaBBC2015,
author = {BBC},
title = {{Migrant crisis: Hungary migrants start walk to border}},
url = {http://www.bbc.com/news/world-europe-34155701},
year = {2015}
}

@misc{BBC2015,
author = {BBC},
booktitle = {BBC},
title = {{UK to accept 20,000 refugees from Syria by 2020}},
url = {http://www.bbc.com/news/uk-34171148},
year = {2015}
}

@misc{Huff2016,
author = {Leivada, Danae},
booktitle = {Huffington Post},
title = {{Migrants And Refugees Stuck In Greece Face Uncertain Future}},
url = {http://www.huffingtonpost.com/entry/refugees-stuck-in-greece{\_}us{\_}56c60cafe4b08ffac127f5dd},
year = {2016}
}

@misc{Opensociety2016,
author = {{Open Society Initiative}},
booktitle = {Open Society Initiative for Europe},
pages = {7},
title = {{Understanding Migration and Asylum in the European Union}},
url = {https://www.opensocietyfoundations.org/explainers/understanding-migration-and-asylum-european-union},
year = {2016}
}

@techreport{Maiani2016,
author = {Maiani, Francesco},
pages = {1--76},
title = {{The Reform of Dublin III Regulation}},
url = {http://www.europarl.europa.eu/committees/en/supporting-analyses-search.html},
year = {2016}
}

@misc{EMN2017,
author = {{European Migration Network}},
title = {{EMN Ad-Hoc Query on immediate family members applying for asylum at the same time}},
url = {http://www.europarl.europa.eu/committees/en/supporting-analyses-search.html},
year = {2017}
}


@Article{Gale1962,
  author    = {Gale, David and Shapley, Lloyd S},
  journal   = {The American Mathematical Monthly},
  title     = {College admissions and the stability of marriage},
  year      = {1962},
  number    = {1},
  pages     = {9--15},
  volume    = {69},
  publisher = {Taylor \& Francis},
}

@article{Roth1982,
  title={The economics of matching: Stability and incentives},
  author={Roth, Alvin E},
  journal={Mathematics of operations research},
  volume={7},
  number={4},
  pages={617--628},
  year={1982},
  publisher={INFORMS}
}

@article{Roth1984,
  title={The evolution of the labor market for medical interns and residents: a case study in game theory},
  author={Roth, Alvin E},
  journal={Journal of political Economy},
  volume={92},
  number={6},
  pages={991--1016},
  year={1984},
  publisher={The University of Chicago Press}
}

@Article{Sonmez1994,
  author    = {S{\"o}nmez, Tayfun},
  journal   = {Economic Design},
  title     = {Strategy-proofness in many-to-one matching problems},
  year      = {1994},
  pages     = {365--380},
  volume    = {1},
  publisher = {Springer},
}

@article{Balinski1999,
  title={A tale of two mechanisms: student placement},
  author={Balinski, Michel and S{\"o}nmez, Tayfun},
  journal={Journal of Economic Theory},
  volume={84},
  number={1},
  pages={73--94},
  year={1999},
  publisher={Elsevier}
}

@Article{Abdulkadiroglu2003,
  author    = {Abdulkadiro{\u{g}}lu, Atila and S{\"o}nmez, Tayfun},
  journal   = {American Economic Review},
  title     = {School choice: A mechanism design approach},
  year      = {2003},
  number    = {3},
  pages     = {729--747},
  volume    = {93},
  publisher = {American Economic Association},
}

@Article{pathak2008leveling,
  author    = {Pathak, Parag A and S{\"o}nmez, Tayfun},
  journal   = {American Economic Review},
  title     = {Leveling the playing field: Sincere and sophisticated players in the Boston mechanism},
  year      = {2008},
  number    = {4},
  pages     = {1636--1652},
  volume    = {98},
  publisher = {American Economic Association},
}

@Article{Echenique2012,
  author    = {Echenique, Federico and Yenmez, M Bumin},
  journal   = {American Economic Review},
  title     = {How to control controlled school choice},
  year      = {2015},
  number    = {8},
  pages     = {2679--2694},
  volume    = {105},
  publisher = {American Economic Association 2014 Broadway, Suite 305, Nashville, TN 37203},
}


@article{UtkuUnver2010,
  title={Dynamic kidney exchange},
  author={{\"U}nver, M Utku},
  journal={The Review of Economic Studies},
  volume={77},
  number={1},
  pages={372--414},
  year={2010},
  publisher={Wiley-Blackwell}
}

@article{moraga2014tradable,
  title={Tradable immigration quotas},
  author={Moraga, Jes{\'u}s Fern{\'a}ndez-Huertas and Rapoport, Hillel},
  journal={Journal of Public Economics},
  volume={115},
  pages={94--108},
  year={2014},
  publisher={Elsevier}
}

@article{jones2018local,
  title={The local refugee match: Aligning refugees’ preferences with the capacities and priorities of localities},
  author={Jones, Will and Teytelboym, Alexander},
  journal={Journal of Refugee Studies},
  volume={31},
  number={2},
  pages={152--178},
  year={2018},
  publisher={Oxford University Press}
}

@article{andersson2020assigning,
  title={Assigning refugees to landlords in Sweden: Efficient, stable, and maximum matchings},
  author={Andersson, Tommy and Ehlers, Lars},
  journal={The Scandinavian Journal of Economics},
  volume={122},
  number={3},
  pages={937--965},
  year={2020},
  publisher={Wiley Online Library}
}

@article{dyntet,
author = {Ahani, Narges and G\"{o}lz, Paul and Procaccia, Ariel D. and Teytelboym, Alexander and Trapp, Andrew C.},
title = {Dynamic Placement in Refugee Resettlement},
journal = {Operations Research},
volume = {72},
number = {3},
pages = {1087-1104},
year = {2024},
doi = {10.1287/opre.2021.0534},

URL = { 
    
        https://doi.org/10.1287/opre.2021.0534

},
eprint = { 
    
        https://doi.org/10.1287/opre.2021.0534

}
,
    abstract = { Boosting Employment of Resettled RefugeesWhether a resettled refugee finds employment in the United States depends in no small part on which host community they are first welcomed to. Every week, resettlement agencies are assigned a group of refugees who they are required to place in communities around the country. In “Dynamic Placement in Refugee Resettlement,” Ahani, Gölz, Procaccia, Teytelboym, and Trapp develop an allocation system that recommends where to place an incoming refugee family with the aim of boosting the overall employment success. Should capacities in high-employment areas be used up as quickly as possible, or does it make sense to hold back for a perfect match? The simple algorithm, based on two-stage stochastic programming, achieves over 98\% of the hindsight-optimal employment, compared with under 90\% for the greedy-like approaches that were previously used in practice. Their algorithm is now part of the Annie™ MOORE optimization software used by a leading American refugee resettlement agency. }
}

@inproceedings{sun2023daycare,
  title={Daycare matching in Japan: Transfers and siblings},
  author={Sun, Zhaohong and Takenami, Yoshihiro and Moriwaki, Daisuke and Tomita, Yoji and Yokoo, Makoto},
  booktitle={Proceedings of the AAAI Conference on Artificial Intelligence},
  volume={37},
  number={12},
  pages={14487--14495},
  year={2023}
}

@Article{Pereyra2013,
  title={A dynamic school choice model},
  author={Pereyra, Juan Sebasti{\'a}n},
  journal={Games and economic behavior},
  volume={80},
  pages={100--114},
  year={2013},
  publisher={Elsevier}
}

@Article{Kurino2014,
  author    = {Kurino, Morimitsu},
  journal   = {American Economic Journal: Microeconomics},
  title     = {House allocation with overlapping generations},
  year      = {2014},
  number    = {1},
  pages     = {258--289},
  volume    = {6},
  publisher = {American Economic Association 2014 Broadway, Suite 305, Nashville, TN 37203-2425},
}


@Article{Aygun2013a,
  author    = {Ayg{\"u}n, Orhan and S{\"o}nmez, Tayfun},
  journal   = {American Economic Review},
  title     = {Matching with contracts: Comment},
  year      = {2013},
  number    = {5},
  pages     = {2050--2051},
  volume    = {103},
  publisher = {American Economic Association},
}

@article{hirata2014cumulative,
  title={Cumulative offer process is order-independent},
  author={Hirata, Daisuke and Kasuya, Yusuke},
  journal={Economics Letters},
  volume={124},
  number={1},
  pages={37--40},
  year={2014},
  publisher={Elsevier}
}


@article{Capari2021,
author = {Caspari, Gian},
year = {2019},
title = {An alternative approach to asylum assignment}
}

@Article{Philipp2015,
  title={Matching with waiting times: The German entry-level labor market for lawyers},
  author={Dimakopoulos, Philipp D and Heller, C-Philipp},
  journal={Games and Economic Behavior},
  volume={115},
  pages={289--313},
  year={2019},
  publisher={Elsevier}
}


@Article{Sonmez2013,
  author    = {S{\"o}nmez, Tayfun and Switzer, Tobias B},
  journal   = {Econometrica},
  title     = {Matching with (branch-of-choice) contracts at the United States Military Academy},
  year      = {2013},
  number    = {2},
  pages     = {451--488},
  volume    = {81},
  publisher = {Wiley Online Library},
}

@inproceedings{Kominers2013,
  title={Designing for diversity in matching},
  author={Kominers, Scott Duke and S{\"o}nmez, Tayfun},
  booktitle={EC},
  pages={603--604},
  year={2013}
}


@article{Kelso1982,
  title={Job matching, coalition formation, and gross substitutes},
  author={Kelso Jr, Alexander S and Crawford, Vincent P},
  journal={Econometrica},
  pages={1483--1504},
  year={1982}
}

@article{HatfieldMilgrom2005,
  title={Matching with contracts},
  author={Hatfield, John William and Milgrom, Paul R},
  journal={American Economic Review},
  volume={95},
  number={4},
  pages={913--935},
  year={2005}

@Article{Hatfield2010,
  author    = {Hatfield, John William and Kojima, Fuhito},
  journal   = {Journal of Economic Theory},
  title     = {Substitutes and stability for matching with contracts},
  year      = {2010},
  number    = {5},
  pages     = {1704--1723},
  volume    = {145},
  publisher = {Elsevier},
}

@article{kominers2016matching,
  title={Matching with slot-specific priorities: Theory},
  author={Kominers, Scott Duke and S{\"o}nmez, Tayfun},
  journal={Theoretical Economics},
  volume={11},
  number={2},
  pages={683--710},
  year={2016},
  publisher={Wiley Online Library}
}

@article{nguyen2021stability,
  title={Stability in matching markets with complex constraints},
  author={Nguyen, Hai and Nguyen, Th{\`a}nh and Teytelboym, Alexander},
  journal={Management Science},
  volume={67},
  number={12},
  pages={7438--7454},
  year={2021},
  publisher={INFORMS}
}

@article{moraga2021can,
  title={Can market mechanisms solve the refugee crisis?},
  author={Moraga, Jes{\'u}s Fern{\'a}ndez-Huertas and Hagen, Martin},
  journal={IZA World of Labor},
  year={2021},
  publisher={Bonn: Institute of Labor Economics (IZA)}
}

@article{troyan2020obvious,
  title={Obvious manipulations},
  author={Troyan, Peter and Morrill, Thayer},
  journal={Journal of Economic Theory},
  volume={185},
  pages={104970},
  year={2020},
  publisher={Elsevier}
}

@article{hagen2024optimal,
  title={Optimal Refugee Status Determination},
  author={Hagen, Martin},
  journal={Available at SSRN 4703869},
  year={2024}
}

@article{hagen2024refugee,
  title={Refugee relocation: A mechanism design approach},
  author={Hagen, Martin},
  journal={The Economic Journal},
  volume={134},
  number={663},
  pages={3027--3046},
  year={2024},
  publisher={Oxford University Press}
}

@article{hagen2022tradable,
  title={Tradable immigration quotas revisited},
  author={Hagen, Martin},
  journal={Journal of Public Economics},
  volume={208},
  pages={104619},
  year={2022},
  publisher={Elsevier}
}

@inproceedings{farajzadeh2023optimizing,
  title={Optimizing Sponsored Humanitarian Parole},
  author={Farajzadeh, Fatemeh and Killea, Ryan Baylor and Teytelboym, Alexander and Trapp, Andrew Christopher},
  booktitle={Proceedings of the 3rd ACM Conference on Equity and Access in Algorithms, Mechanisms, and Optimization},
  pages={1--13},
  year={2023}
}

@article{romm2024stability,
  title={Stability vs. no justified envy},
  author={Romm, Assaf and Roth, Alvin E and Shorrer, Ran I},
  journal={Games and Economic Behavior},
  volume={148},
  pages={357--366},
  year={2024},
  publisher={Elsevier}
}

@article{hatfield2016hidden,
  title={Hidden substitutes},
  author={Hatfield, John William and Kominers, Scott Duke},
  year={2019},
  publisher={Working Paper, Harvard University}
}

Hatfield, J. W. and S. D. Kominers (2019). Hidden substitutes. Working paper, Harvard
University

@article{hatfield2017stability,
    author = {Hatfield, John William and Kominers, Scott Duke and Westkamp, Alexander},
    title = "{Stability, Strategy-Proofness, and Cumulative Offer Mechanisms}",
    journal = {The Review of Economic Studies},
    volume = {88},
    number = {3},
    pages = {1457-1502},
    year = {2020},
    month = {09},
    abstract = "{We characterize when a stable and strategy-proof mechanism is guaranteed to exist in the setting of many-to-one matching with contracts. We introduce three novel conditions—observable substitutability, observable size monotonicity, and non-manipulability via contractual terms—and show that when these conditions are satisfied, the cumulative offer mechanism is the unique mechanism that is stable and strategy-proof (for workers). Moreover, we show that our three conditions are, in a sense, necessary: if the choice function of some firm fails any of our three conditions, we can construct unit-demand choice functions for the other firms such that no stable and strategy-proof mechanism exists. Thus, our results provide a rationale for the ubiquity of cumulative offer mechanisms in practice.}",
    issn = {0034-6527},
    doi = {10.1093/restud/rdaa052},
    url = {https://doi.org/10.1093/restud/rdaa052},
    eprint = {https://academic.oup.com/restud/article-pdf/88/3/1457/38107855/rdaa052.pdf},
}


@Article{Utku2015,
  author  = {Dur, Umut and {\"U}nver, M Utku},
  journal = {Available at SSRN 2180357},
  title   = {Two-sided matching via balanced exchange: Tuition and worker exchanges},
  year    = {2015},
}

@TechReport{kotowski2015note,
  author    = {Kotowski, Maciej H and others},
  title     = {A note on stability in one-to-one, multi-period matching markets},
  year      = {2015},
  publisher = {Harvard Univ., John F. Kennedy School of Government},
}


@Article{Roth2004,
  author    = {Roth, Alvin E and S{\"o}nmez, Tayfun and {\"U}nver, M Utku},
  journal   = {The Quarterly journal of economics},
  title     = {Kidney exchange},
  year      = {2004},
  number    = {2},
  pages     = {457--488},
  volume    = {119},
  publisher = {MIT Press},
}


@Article{hassidim2017redesigning,
  author    = {Hassidim, Avinatan and Romm, Assaf and Shorrer, Ran I},
  title     = {Redesigning the Israeli psychology master's match},
  year      = {2017},
  pages     = {121--122},
  booktitle = {Proceedings of the 2017 ACM Conference on Economics and Computation},
}

@Article{aygun2016dynamic,
  title={Dynamic reserves in matching markets},
  author={Ayg{\"u}n, Orhan and Turhan, Bertan},
  journal={Journal of Economic Theory},
  volume={188},
  pages={105069},
  year={2020},
  publisher={Elsevier}
}

@article{yenmez2018college,
  title={A college admissions clearinghouse},
  author={Yenmez, M Bumin},
  journal={Journal of Economic Theory},
  volume={176},
  pages={859--885},
  year={2018},
  publisher={Elsevier}
}


@Article{mcdermid2010keeping,
  author    = {McDermid, Eric J and Manlove, David F},
  journal   = {Journal of Combinatorial Optimization},
  title     = {Keeping partners together: algorithmic results for the hospitals/residents problem with couples},
  year      = {2010},
  pages     = {279--303},
  volume    = {19},
  publisher = {Springer},
}

@article{biro2014matching,
  title={Matching with sizes (or scheduling with processing set restrictions)},
  author={Bir{\'o}, P{\'e}ter and McDermid, Eric},
  journal={Discrete Applied Mathematics},
  volume={164},
  pages={61--67},
  year={2014},
  publisher={Elsevier}
}

@InProceedings{dean2006unsplittable,
  author       = {Dean, Brian C and Goemans, Michel X and Immorlica, Nicole},
  booktitle    = {Fourth IFIP International Conference on Theoretical Computer Science-TCS 2006: IFIP 19th Worm Computer Congress, TC-1, Foundations of Computer Science, August 23--24, 2006, Santiago, Chile},
  title        = {The unsplittable stable marriage problem},
  year         = {2006},
  organization = {Springer},
  pages        = {65--75},
}

@article{yenmez2018college,
  title={A college admissions clearinghouse},
  author={Yenmez, M Bumin},
  journal={Journal of Economic Theory},
  volume={176},
  pages={859--885},
  year={2018},
  publisher={Elsevier}
}

@Article{nguyen2018near,
  author    = {Nguyen, Thanh and Vohra, Rakesh},
  journal   = {American Economic Review},
  title     = {Near-feasible stable matchings with couples},
  year      = {2018},
  number    = {11},
  pages     = {3154--3169},
  volume    = {108},
  publisher = {American Economic Association 2014 Broadway, Suite 305, Nashville, TN 37203},
}

@article{delacretaz2019stability,
  title={Stability in matching markets with sizes},
  author={Delacr{\'e}taz, David},
  year={2019},
  publisher={Working Paper}
}


@article{roth1997effects,
  title={The effects of the change in the NRMP matching algorithm},
  author={Roth, Alvin E and Peranson, Elliott},
  journal={JAMA},
  volume={278},
  number={9},
  pages={729--732},
  year={1997},
  publisher={American Medical Association}
}

@Article{roth1999redesign,
  author    = {Roth, Alvin E and Peranson, Elliott},
  journal   = {American economic review},
  title     = {The redesign of the matching market for American physicians: Some engineering aspects of economic design},
  year      = {1999},
  number    = {4},
  pages     = {748--780},
  volume    = {89},
  publisher = {American Economic Association},
}

@article{roth2003origins,
  title={The origins, history, and design of the resident match},
  author={Roth, Alvin E},
  journal={Jama},
  volume={289},
  number={7},
  pages={909--912},
  year={2003},
  publisher={American Medical Association}
}


@Article{kennes2014day,
  author    = {Kennes, John and Monte, Daniel and Tumennasan, Norovsambuu},
  journal   = {American Economic Journal: Microeconomics},
  title     = {The day care assignment: A dynamic matching problem},
  year      = {2014},
  number    = {4},
  pages     = {362--406},
  volume    = {6},
  publisher = {American Economic Association 2014 Broadway, Suite 305, Nashville, TN 37203-2425},
}

@article{kennes2011daycare,
  title={The Daycare Assignment Problem: Matching in an Overlapping Generations Model},
  author={Kennes, John and Monte, Daniel and Tumennasan, Norovsambuu},
  year={2011}
}


@Article{doval2019dynamically,
  author    = {Doval, Laura},
  journal   = {Theoretical Economics},
  title     = {Dynamically stable matching},
  year      = {2022},
  number    = {2},
  pages     = {687--724},
  volume    = {17},
  publisher = {Wiley Online Library},
}


@article{hirata2014cumulative,
  title={Cumulative offer process is order-independent},
  author={Hirata, Daisuke and Kasuya, Yusuke},
  journal={Economics Letters},
  volume={124},
  number={1},
  pages={37--40},
  year={2014},
  publisher={Elsevier}
}

@Comment{jabref-meta: databaseType:bibtex;}